\documentclass[12pt,a4paper]{article}

\usepackage[dvipsnames]{xcolor}
\usepackage{jheppub}

\usepackage{amsmath}
\usepackage{amssymb}
\usepackage{amsfonts}
\usepackage{amsthm}
\usepackage{mathtools}
\usepackage{mathrsfs}
\usepackage{dsfont}
\usepackage{braket}

\usepackage{tikz}
\usetikzlibrary{decorations.pathreplacing, decorations.markings, calc, shapes.misc, decorations.pathmorphing, patterns.meta, math, shapes.geometric}
\tikzset{snake it/.style={decorate, decoration=snake}}
\tikzset{cross/.style={cross out, draw=black, minimum size=2*(#1-\pgflinewidth), inner sep=0pt, outer sep=0pt},
cross/.default={1pt}}
\tikzset{
    partial ellipse/.style args={#1:#2:#3}{
        insert path={+ (#1:#3) arc (#1:#2:#3)}
    }
}
\usepackage{subcaption}
\usepackage[most]{tcolorbox}

\hypersetup{
    pdfencoding=unicode,
    colorlinks=true,
    urlcolor=Maroon,
    linkcolor=RoyalBlue,
    citecolor=Maroon,
    pdftitle={The Runkel-Watts string},
    pdfauthor={Mattia Biancotto, Lorenz Eberhardt, Victor A. Rodriguez, Zi-Yue Wang},
    pdfdisplaydoctitle=true,
    pdfstartview=FitH,
    linktocpage=true
}

\newcommand{\ba}{\begin{align}}

\newcommand{\be}{\begin{equation}}
\newcommand{\ee}{\end{equation}}
\def\bd{\begin{tikzpicture}}
\def\ed{\end{tikzpicture}}

\DeclareMathOperator\tr{tr}
\newcommand\e{\mathop{\text{e}}}
\newcommand\ex{{\mathrm e}}

\renewcommand\Re{\mathop{\text{Re}}}
\newcommand\Res{\mathop{\text{Res}}}
\DeclareMathOperator{\Aut}{Aut}

\DeclareMathOperator{\sgn}{\text{sgn}}

\renewcommand\d{\text{d}}

\allowdisplaybreaks[1]

\newcommand\SU{\text{SU}}

\newcommand\CC{\mathbb{C}}
\newcommand\ZZ{\mathbb{Z}}
\newcommand\RR{\mathbb{R}}

\newcommand{\id}{\mathds{1}}

\newcommand{\bM}{\overline{\mathcal{M}}}

\newcommand{\xx}{\mathsf{x}}
\newcommand{\yy}{\mathsf{y}}
\newcommand{\CLS}{\ensuremath{\mathbb{C}\mathrm{LS}}}

\def \bal#1\eal  {\begin{align} #1 \end{align}}
\newcommand{\beq} {\begin{equation}}
\newcommand{\eeq} {\end{equation}}
\newcommand{\nn} {\nonumber\\~}

\title{The Runkel-Watts string}

\author{Mattia Biancotto$^a$,}
\author{Lorenz Eberhardt$^b$,}
\author{Victor A. Rodriguez$^c$,}
\author{Zi-Yue Wang$^c$}

\affiliation[a]{Max-Planck-Institut f\"ur Gravitationsphysik (Albert-Einstein-Institut)
Am M\"uhlenberg 1, 14476 Potsdam, Germany}

\affiliation[b]{Institute for Theoretical Physics,
University of Amsterdam, Amsterdam, 1098XH, NL}
\affiliation[c]{Department of Physics, University of California, Santa Barbara, CA 93106, USA}

\emailAdd{mattia.biancotto@aei.mpg.de}
\emailAdd{l.eberhardt@uva.nl}
\emailAdd{varodriguez@ucsb.edu}
\emailAdd{zi-yue@ucsb.edu}

\abstract{
    We introduce and solve a new family of two-dimensional string theories obtained by coupling Liouville CFT to the generalized Runkel-Watts CFT. 
    By taking an appropriate limit of the complex Liouville string, we derive an all-genus formula for its string amplitudes, and formulate a duality with a matrix integral.
    We explain that it can be interpreted as 2d $\mathrm{SU}(2)$ Yang-Mills theory coupled to gravity.
    We show that it also directly relates to other minimal string constructions such as the A-series minimal string.
}
\begin{document}

\maketitle

\makeatletter
\g@addto@macro\bfseries{\boldmath}
\makeatother

\section{Introduction}\label{sec:introduction}

Low-dimensional string theories such as the minimal string \cite{Brezin:1990rb,Douglas:1989ve,Gross:1989vs,Seiberg:2004at} and the $c=1$ string \cite{Klebanov:1991qa,Ginsparg:1993is,Jevicki:1993qn,Polchinski:1994mb,Balthazar:2019rnh} have long served as valuable laboratories for quantum gravity and holography. 
In recent years, this subject has entered a phase in which the construction, solution, and interconnection of such models have come under increasingly precise control. 
A new family is the Virasoro minimal string (VMS) \cite{Collier:2023cyw}, which provides a continuous version of ordinary minimal string theory and, in particular, gives a stringy realization and generalization of the duality between JT gravity and random matrix theory \cite{Saad:2019lba}. 
A second family is the complex Liouville string (\CLS{}), a substantially richer model whose observables are nevertheless under good computational control \cite{Collier:2024kmo,Collier:2024kwt,Collier:2024lys,Collier:2024mlg}. 
Notably, the \CLS{} admits several interpretations: as a model of two-dimensional strings, as a two-dimensional sine-dilaton theory of gravity \cite{Collier:2025pbm}, and as a source of integrated cosmological correlators in three-dimensional de Sitter gravity \cite{Collier:2025lux}.

Despite the complexity of their worldsheet observables, these theories have turned out to be surprisingly soluble. 
Their string amplitudes can be studied using several a priori distinct approaches: exact worldsheet conformal field theory (CFT), analytic bootstrap from properties of Liouville correlation functions, the geometry of the moduli space of Riemann surfaces, and topological recursion of matrix integrals. 

The different theories in this landscape are also not isolated from one another. 
Most notably, the non-compact $c=1$ string was recently shown to fit naturally into the same matrix integral and intersection-theoretic framework \cite{Collier:2026pxi}. 
This result connects two seemingly distinct corners of the two-dimensional string landscape: on the one hand, minimal string theories, whose prototypical matter sectors are rational Virasoro minimal models and which are usually interpreted as models of two-dimensional quantum gravity on the worldsheet; and on the other hand, the $c=1$ string, whose matter sector is a non-compact timelike free boson and whose observables admit a more direct interpretation as target-space $S$-matrix elements.

More broadly, a pattern begins to emerge.
The complex Liouville string appears to sit near the top of a hierarchy of 2d string theories, with several simpler models arising from it by appropriate limits.\footnote{There are, of course, other bosonic 2d string theories that do not fit into this pattern, such as ADE minimal strings and $c=1$ orbifolds. These are not presently known to arise as limits of the \CLS{}.}
At the same time, the Virasoro minimal string appears as a universal skeleton inside these limits: the VMS quantum volumes provide the elementary building blocks from which more complicated string amplitudes are assembled. 

\paragraph{A new family of two-dimensional strings.}
The main purpose of this paper is to introduce and solve a new family of two-dimensional string theories that follows this pattern. 
The worldsheet theory is obtained by coupling Liouville CFT to the Runkel-Watts CFT\footnote{The Runkel-Watts CFT was originally discovered at central charge $c=1$ \cite{Runkel:2001ng}, and was later generalized to other rational central charges with $c\leq 1$ \cite{Schomerus:2003vv,McElgin:2007ak}. In this paper we will refer to this entire family as Runkel-Watts CFT. It is also often called non-analytic Liouville CFT.}, together with the usual $\mathfrak{b}\mathfrak{c}$-ghost system,
\begin{equation}
\begin{array}{c}
\text{Runkel-Watts CFT} \\ \text{$26 - c\leq 1$}
\end{array}
\ \otimes\  
\begin{array}{c} \text{Liouville CFT} \\ \text{$c \geq 25$} \end{array} \ \otimes\  
\begin{array}{c} \text{$\mathfrak{b}\mathfrak{c}$-ghosts} \\ \text{$c= -26$} \end{array}\,.
\label{eq:RW string intro}
\end{equation}
and we will refer to the resulting critical string theory as the \emph{Runkel-Watts string}. 
Like the underlying CFT, the Runkel-Watts string is a discrete family of string theories labeled by two coprime integers $(q,q')$. 
The Runkel-Watts CFT is a close cousin of timelike Liouville CFT, but its three-point structure constant is multiplied by a chamber function, i.e. a piecewise constant function. 
This makes the theory piecewise analytic as a function of the external Liouville momenta. 
The Runkel-Watts string therefore provides a natural intermediate object between the analytic world of Liouville-type strings and the rational world of ordinary minimal strings.

\begin{figure}
    \centering
    \begin{tikzpicture}
        \node[rectangle, rounded corners, draw, very thick, fill=blue!10!white, text width=8em, text centered, outer sep=3pt] (CLS) at (0,4) {Complex Liouville \\ string};
        \node[rectangle, rounded corners, draw, very thick, fill=blue!10!white, text width=8em, text centered, outer sep=3pt] (c1) at (3.5,2) {non-compact \\ $c=1$ string};
        \node[rectangle, rounded corners, draw, very thick, fill=blue!10!white, text width=8em, text centered, outer sep=3pt] (RWS) at (-3.5,2) {Runkel-Watts \\ string};
        \node[rectangle, rounded corners, draw, very thick, fill=blue!10!white, text width=8em, text centered, outer sep=3pt] (VMS) at (-3.5,0) {Virasoro \\ minimal string};
        \node[rectangle, rounded corners, draw, very thick, fill=blue!10!white, text width=8em, text centered, outer sep=3pt] (MS) at (-7,4) {A-series \\ minimal string};
        \draw[very thick,->] (CLS) to node[shift={(-.5,.2)}] {\ref{subsec:amplitudes-from-geometry}} (RWS);
        \draw[very thick,->] (MS) to node[shift={(.5,.2)}] {\ref{subsec:A-series-to-RW-CFT}} (RWS);
        \draw[very thick,->] (RWS) to node[shift={(.5,0)}] {\ref{subsec:factorized-VMS}} (VMS);
        \draw[very thick,->] (CLS) to node[shift={(.5,.2)}] {\cite{Collier:2026pxi}} (c1);
    \end{tikzpicture}
    \caption{A schematic view of the relations among several 2d string theories. The Runkel-Watts string sits at an intermediate point between the complex Liouville string, the A-series minimal string, and a coarse-grained version of the Virasoro minimal string.}
    \label{fig:minimal string theories}
\end{figure}
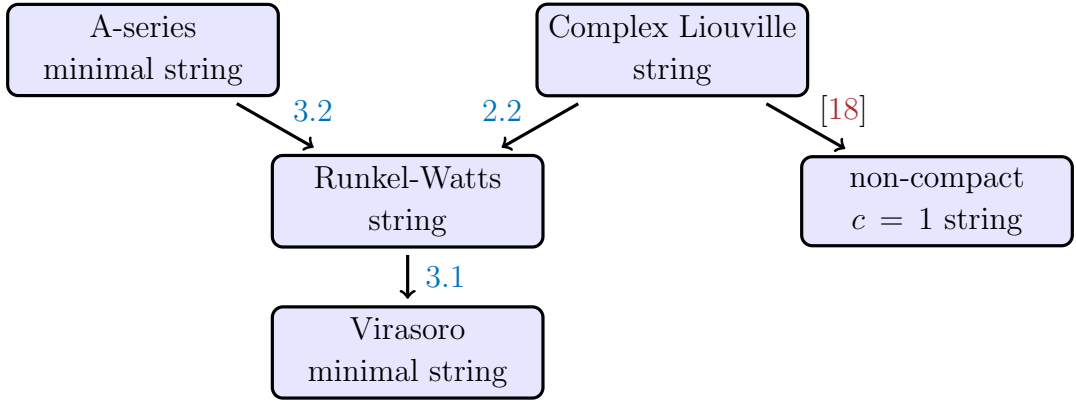

This intermediate position is one of the main conceptual points of the paper and is summarized in figure~\ref{fig:minimal string theories}.
On the one hand, the Runkel-Watts string can be obtained from the \CLS{} by a limit in which one of its Liouville factors degenerates to the Runkel-Watts CFT. 
On the other hand, it is related to the A-series minimal string through the corresponding limiting relation between their matter CFTs. 
Finally, in a suitable large-$(q,q')$ limit, its amplitudes factorize into a two-dimensional gauge theory factor and a VMS quantum volume, as described below.

\paragraph{Feynman rules for string amplitudes.}
Our main computational result is an all-genus expression for the Runkel-Watts string amplitudes.
A striking feature of the \CLS{} and the $c=1$ string is that their amplitudes admit stable graph expansions that may be interpreted as Feynman rules \cite{Collier:2024lys,Collier:2026pxi}, akin to those of closed string field theory \cite{Sen:2024nfd}.
The VMS quantum volumes assigned to the vertices provide the universal gravitational building blocks, while the edge, leg, and color data encode their theory-dependent gluing.

We find an analogous expression for the Runkel-Watts string:
\begin{align}\label{eq:RW Feynman rules intro}
    \mathsf{RW}_{g,n}(\boldsymbol{p})
    = \sum_{\Gamma \in \mathcal{G}_{g,n}} \frac{1}{|\text{Aut}(\Gamma)|} &\int' \prod_{e \in \mathcal{E}_\Gamma}(-2 |k_e|\d k_e)
    \nonumber\\
    &\times \prod_{v \in \mathcal{V}_\Gamma}
    \mathsf{A}_{g_v,n_v}^\text{TQFT}\big(\sqrt{qq'}\boldsymbol{p}_v+\tfrac{1}{2}(q+q')\big) \,
    \mathsf{V}_{g_v,n_v}^{(\sqrt{q'/q})}(i \boldsymbol{p}_v) \, .
\end{align}
Here, $\boldsymbol{p}=(p_1,\ldots,p_n)$, and $\mathcal{G}_{g,n}$ is the set of stable graphs representing stable degenerations of the worldsheet Riemann surface.
For $\Gamma\in\mathcal{G}_{g,n}$, $\mathcal{E}_\Gamma$ and $\mathcal{V}_\Gamma$ denote its edge and vertex sets.
Each internal edge carries an integrated momentum $k_e$, with the prime on the integral denoting the appropriate regularization.
For each vertex $v$, $g_v$ and $n_v$ denote its genus and valence, while $\boldsymbol{p}_v$ is the list of external and internal momenta incident on it.
The contribution of each vertex is the product of a partition function $\mathsf{A}_{g_v,n_v}^\text{TQFT}$ of a two-dimensional topological quantum field theory (TQFT) and a VMS quantum volume $\mathsf{V}_{g_v,n_v}$.
For the Runkel-Watts string, this TQFT is two-dimensional $\mathrm{SU}(2)$ Yang-Mills theory.
Since the VMS admits an interpretation as a theory of quantum gravity, \eqref{eq:RW Feynman rules intro} may therefore be viewed as a particular coupling of two-dimensional $\mathrm{SU}(2)$ Yang-Mills theory to gravity.
The \CLS{} and $c=1$ string amplitudes have the same structure, with their TQFT data supplied by the $q$-deformed $\mathrm{SU}(2)$ and $\mathrm{U}(1)$ Yang-Mills TQFTs, respectively.

We test \eqref{eq:RW Feynman rules intro} in the first nontrivial cases, $\mathsf{RW}_{1,1}$ and $\mathsf{RW}_{0,4}$, by comparing its predictions with direct numerical integrations of the corresponding worldsheet CFT correlators over moduli space.
We also verify a triality symmetry of the sphere four-point amplitude.
Finally, we derive the dilaton equation from the topological recursion governing the dual matrix integral.
This equation relates a momentum derivative of an amplitude, evaluated at the special dilaton momenta, to the corresponding amplitude with one fewer external leg.

\paragraph{A matrix integral for the Runkel-Watts string.}
The same amplitudes can also be generated by topological recursion \cite{Eynard:2007kz}, leading to a new two-dimensional string/matrix-integral duality.
We derive this duality by exploiting the realization of the Runkel-Watts structure constants as a limit of Liouville structure constants, and by taking the corresponding limit of the known \CLS{} duality.
This limit differs from the one that produces the non-compact $c=1$ string, but the resulting structure closely parallels the $c=1$ case.\footnote{In view of this similarity, it would be interesting to understand whether the Runkel-Watts string amplitudes admit an interpretation as asymptotic observables in some target-space description.}
The dual spectral curve can be parametrized as
\begin{equation}\label{eq:RW spectral curve intro}
    \xx(z)
    =
    -2\cos\!\left(\frac{\pi\sqrt{z}}{\sqrt{qq'}}\right)\,,
    \qquad
    \yy(z)
    =
    -4\pi\sqrt{z}\,
    \frac{
        \sin\!\left(\frac{\pi q'\sqrt{z}}{\sqrt{qq'}}\right)
        \sin\!\left(\frac{\pi q\sqrt{z}}{\sqrt{qq'}}\right)
    }{
        \sin\!\left(\frac{\pi\sqrt{z}}{\sqrt{qq'}}\right)
    }\,,
\end{equation}
where $z\in\mathbb{C}$ is a coordinate on the spectral curve.
This curve determines the perturbative expansion of the matrix integral through topological recursion.
As in the \CLS{} case, the spectral curve is non-algebraic and has infinitely many sheets and branch points satisfying $\d\xx(z)=0$.
The resulting topological-recursion solution belongs to the class of solutions of the loop equations for solvable two-matrix models with the standard bilinear coupling $\tr M_1M_2$ \cite{Chekhov:2006vd}.

\paragraph{Consequences of the limiting relations.}
The limiting relations with other minimal strings have interesting consequences.
The relation to the A-series minimal string may help identify the intersection-theoretic formula and Feynman rules for general A-series amplitudes (see \cite{Artemev:2025pvk} for the $(2,p)$ case).

The Runkel-Watts string also admits a factorized gauge$\times$gravity limit.
In a large-$(q,q')$ limit, the edge contributions in \eqref{eq:RW Feynman rules intro} are suppressed, so that the single-vertex stable graph dominates.
The amplitudes consequently factorize into the corresponding VMS quantum volume and the two-dimensional gauge theory factor $\mathsf{A}^{\mathrm{TQFT}}$; the latter oscillates increasingly rapidly as a function of the external momenta, while the VMS quantum volume determines the envelope of the full amplitude.

\paragraph{Outline of the paper.}
This paper is organized as follows. In section~\ref{sec:RW-string-from-CLS} we construct the Runkel-Watts string from the \CLS{} and solve its perturbative amplitudes. We begin in section~\ref{subsec:worldsheet-CFT} by reviewing the relevant Liouville and Runkel-Watts CFT data and fixing the string normalizations. In section~\ref{subsec:amplitudes-from-geometry} we take the rational Runkel-Watts limit of the \CLS{} intersection-number formula, derive the stable graph expansion for $\mathsf{RW}_{g,n}$, and rewrite it as momentum-space Feynman rules with $\mathrm{SU}(2)$ Yang-Mills TQFT factors and VMS quantum volumes at the vertices. In section~\ref{subsec:examples and checks} we evaluate $\mathsf{RW}_{1,1}$ and $\mathsf{RW}_{0,4}$ explicitly and check them against triality and direct numerical integration. In section~\ref{subsec:properties} we record general structural properties of the amplitudes, including piecewise polynomiality, discontinuities without poles, polynomial growth, and the dilaton equation. In section~\ref{subsec:dual-spectral-curve} we extract the dual spectral curve, formulate the associated topological recursion, describe the dictionary between resolvents and Runkel-Watts amplitudes, and recover the low-point amplitudes from this recursion.
In section~\ref{sec:relation-to-other-minimal-strings} we explain how the Runkel-Watts string fits into the broader network of minimal strings. Section~\ref{subsec:factorized-VMS} studies the factorized gauge$\times$gravity limit of the amplitudes, while section~\ref{subsec:A-series-to-RW-CFT} derives the Runkel-Watts CFT, and then the Runkel-Watts string, as a limit of A-series minimal models and minimal strings. Appendix~\ref{app:details-of-calculations} contains the details of several calculations used in the main text.

%%%%%%%%%%%%%%%%%%%%%%%%%%%%%%%%%%%%%%%%%%%%%%%%%%%%%%%%%%%%%%%%
\section{The Runkel-Watts string from \texorpdfstring{$\mathbb{C}$LS}{CLS}}\label{sec:RW-string-from-CLS}

\subsection{Worldsheet CFT}\label{subsec:worldsheet-CFT}

\paragraph{Liouville CFT.}
We begin by briefly recalling the conventions that we use for Liouville CFT, following in particular the normalization of \cite{Collier:2024kwt}. Liouville theory is a non-compact solution of the conformal bootstrap with central charge
\be
    c=1+6Q^2\,.
\ee
Here $Q=b+b^{-1}$. For $c>1$, its normalizable spectrum consists of a continuum of scalar Virasoro primaries $V_p$ whose conformal weights are parameterized as
\be
    h_p=\tilde h_p=\frac{Q^2}{4}-p^2\,,\qquad p\in i\RR_{\geq 0}\,.
\ee
The identity operator is obtained by analytic continuation to $p\to\frac{Q}{2}$. The local CFT data are fixed by the DOZZ structure constants \cite{Dorn:1994xn,Zamolodchikov:1995aa,Teschner:1995yf}, which in our reflection-symmetric normalization take the form
\be\label{eq:DOZZ}
    C_b(p_1,p_2,p_3)=
    \frac{\Gamma_b(2Q)\Gamma_b(\frac{Q}{2}\pm p_1\pm p_2\pm p_3)}
    {\sqrt{2}\Gamma_b(Q)^3\prod_{j=1}^3 \Gamma_b(Q\pm 2p_j)}\,,
\ee
where the $\pm$ notation denotes the product over all independent choices of signs. The corresponding two-point function is obtained by setting one operator to the identity and gives the OPE measure
\be\label{eq:Liouville OPE measure}
    \rho_b(p)=4\sqrt{2}\sin(2\pi bp)\sin(2\pi b^{-1}p)\,.
\ee
For a summary of the analytic properties of the Barnes double gamma function $\Gamma_b$, see for instance \cite{Eberhardt:2023mrq}.
Together with Virasoro conformal blocks, this data determines Liouville correlators on arbitrary closed Riemann surfaces by sewing pairs of pants and integrating the internal Liouville momenta along the spectral contour. Although the physical spectrum lies on $p\in i\RR_{\geq 0}$, the structure constants are meromorphic functions of the external momenta, and Liouville correlators may be analytically continued in the $p_j$'s, up to the familiar contour-deformation subtleties that occur when poles of the DOZZ factors cross the internal OPE contours.

\paragraph{Runkel-Watts CFT.}
Runkel-Watts CFT is the $c=1$ prototype of a family of non-analytic Liouville theories at rational central charge \cite{Runkel:2001ng}. Following the generalization of \cite{McElgin:2007ak}, motivated in part by the $c=1$ limit studied in \cite{Schomerus:2003vv}, we will refer to the whole family as Runkel-Watts CFT, or RW-CFT, with
\be
c=1-6(\beta - \beta^{-1})^2\,,\qquad \beta^2  = \frac{q'}{q}\,,\qquad q,q' \in \ZZ_{>0} \text{ coprime}\,.
\ee
We use upper-case momenta for this theory. Its spectrum is diagonal and continuous, with scalar Virasoro primaries $\widehat{V}_P$ labeled by $P\in\RR_{\geq 0}$, and with conformal weights
\be
    h_P=\tilde h_P=\frac{c-1}{24}+P^2\,.
\ee
 
\begin{figure}[h]
    \centering
    \includegraphics[width=0.6\textwidth]{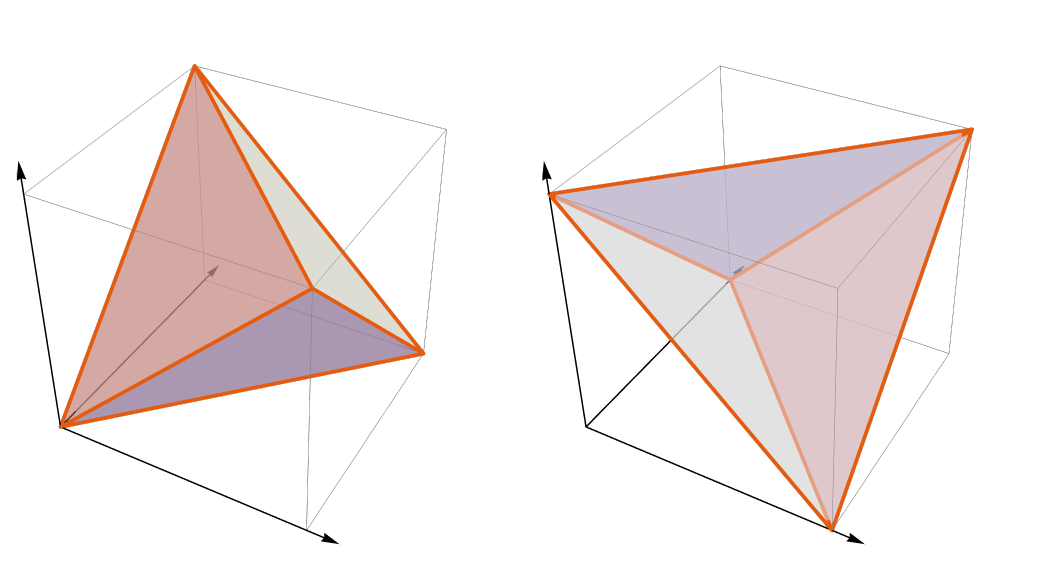}
    \caption{The colored tetrahedron shows the region in which $\sigma(P_1,P_2,P_3)=1$ within the fundamental cube $0\leq P_i\leq \frac{1}{2\sqrt{qq'}}$ for $i=1,2,3$. Left: $q+q'$ even. Right: $q+q'$ odd.}
    \label{fig:tetrahedron}
\end{figure}

The two-point function takes the form
\begin{equation} \label{eq:RW two-point function}
\langle \widehat{V}_{P_1}(0) \widehat{V}_{P_2}(1) \rangle = \frac{\rho_\beta(P_1)}{(P_1)^2} \left( \delta(P_1-P_2) + \delta(P_1+P_2) \right)\,,
\end{equation}
where $\rho_\beta(P)$ is obtained from $\rho_b(p)$ in \eqref{eq:Liouville OPE measure} by setting $b=\beta$ and $p=P$. This is the same as in $c\leq 1$ Liouville CFT, which is sometimes also referred to as timelike Liouville CFT \cite{Ribault:2015sxa,Collier:2023cyw}. The three-point function is given by 
\be\label{eq:RW three-point structure constant}
\langle \widehat{V}_{P_1}(0) \widehat{V}_{P_2}(1) \widehat{V}_{P_3}'(\infty) \rangle = \big( C_\beta(P_1,P_2,P_3) \big)^{-1}\sigma(P_1,P_2,P_3)\,.
\ee
Here the first factor is the inverse of the DOZZ structure constant \eqref{eq:DOZZ}, with the convention that an upper-case argument means that we plug in $p_j=P_j$ in the lower-case Liouville formula. The second factor is the non-analytic step function
\be\label{eq:RW sigma definition}
\sigma(P_1,P_2,P_3) = \begin{cases}
1& \text{if } \prod_{\pm, \pm}\sin \big(\frac{q+q'}{2}\pi+\sqrt{qq'}(P_1\pm P_2\pm P_3)\pi\big)<0\,,\\
0& \text{otherwise}
\end{cases}\,.
\ee
This definition\footnote{In these periodic factors, the shift $\frac{q+q'}{2}$ could equally be written as $\frac{q-q'}{2}$ or $\frac{q'-q}{2}$: the differences are integers. We use the $q+q'$ form for symmetry.} is invariant under permutations of the $P_j$'s and under independent reflections $P_j\to -P_j$: flips of $P_2$ or $P_3$ simply relabel the signs, while the flip of $P_1$ leaves the product unchanged because it replaces the four sine arguments by their reflected partners. Moreover, $\sigma$ is periodic in each argument independently, with period $1/\sqrt{qq'}$. The role of $\sigma$ is to restrict the allowed regions of momentum space; see figure~\ref{fig:tetrahedron}. In particular,\footnote{For example, if $P_1=m/(2\sqrt{qq'})$, the four sine factors pair up into a square, up to the common shift by an integer or half-integer multiple of $\pi$, so their product cannot be negative. The other cases follow by permutation symmetry.}
\be\label{eq:RW sigma lattice zero}
\sigma(P_1,P_2,P_3)=0
\quad \text{if} \quad
P_i \in \frac{1}{2\sqrt{qq'}}\ZZ
\quad \text{for some } i=1,2,3\,.
\ee
This lattice contains the set of inverse-DOZZ poles that lies along the real $P$-line. Indeed, the poles of the inverse-DOZZ structure constant in \eqref{eq:RW three-point structure constant} occur at
\be
    \pm P_j=\frac{m+1}{2}\beta+\frac{n+1}{2}\beta^{-1}\,,\qquad m,n\in\ZZ_{\geq 0}\,.
\ee
After setting $\beta=\sqrt{q'/q}$, these loci lie at $P_j\in \frac{1}{2\sqrt{qq'}}\ZZ$, and are therefore killed by the $\sigma$ factor. Consequently, in any pair-of-pants decomposition, the integration contours over intermediate Runkel-Watts momenta can be kept on $P\in \RR_{\geq 0}$. This contrasts with timelike Liouville CFT, where one shifts the contour away from the real line to avoid these poles \cite{Ribault:2015sxa}.

The resulting theory solves the bootstrap equations, as supported by analytic and numerical checks of crossing symmetry \cite{Ribault:2015sxa}.

\paragraph{Liouville to Runkel-Watts CFT.}

The naive analytic continuation of the spacelike Liouville DOZZ structure constant to $c\leq 1$ is singular in general. However, when the central charge approaches a rational value on this half-line, a finite limit can be extracted \cite{Schomerus:2003vv,McElgin:2007ak,Ribault:2015sxa}. In the normalization conventions of this paper,
\bal \label{eq:Liouville to RW limit}
&\lim_{\varepsilon\rightarrow 0} i\varepsilon q \, C_{i\beta}(iP_1,iP_2,iP_3) \Big|_{\beta=\sqrt{\frac{q'}{q}} + i\varepsilon} \nonumber\\
&\hspace{1cm}= A \prod_{j=1}^3 f(P_j) \times \big ( C_\beta(P_1,P_2,P_3) \big )^{-1} \sigma(P_1,P_2,P_3) \Big |_{\beta=\sqrt{\frac{q'}{q}}} \, .
\eal
Here $\varepsilon>0$ regulates the approach to the rational point from the upper half-plane in $\beta$.
After stripping off the one-point factors $f(P_j)$ and the overall constant $A$, the remaining three-point coupling is precisely the RW-CFT structure constant.

We will derive this relation below and give the explicit form of the factors $f(P_j)$ using the solved sphere three-point amplitude in complex Liouville string theory.
Note that these $f(P_j)$ and $A$ factors can be absorbed into a redefinition of the RW-CFT vertex operators and the Euler counterterm of the CFT, and hence do not affect the CFT data of the RW-CFT, but we will keep the conventions in the previous paragraph. 
Nevertheless, for the purposes of defining the Runkel-Watts string, these factors may also be absorbed into the normalization of the vertex operators $\mathcal{N}_{p}$ together with the normalization constants $C_{\Sigma_g}$ associated with the string path integral on a genus-$g$ Riemann surface $\Sigma_g$. As we will describe shortly, we will fix these in order to obtain a simple relation between the Runkel-Watts string amplitudes and the complex Liouville string amplitudes, which will be our main tool for computing the former.
For completeness, the explicit form of the factors $A$ and $f(P_j)$ is given by
\begin{subequations}
\begin{align} 
    A &= \frac{8\sqrt{2}}{\pi} \left(\frac{q^2-q'^2}{q q'}\right)^2 \sin(\pi \tfrac{q'}{q})\sin(\pi \tfrac{q}{q'}) \,, \\
    f(P) &= \frac{P}{\rho_\beta(P)} \frac{q q'}{q^2-q'^2}\sgn \sin\big(2\pi(\sqrt{qq'}P+\tfrac{q+q'}{2})\big) \,.\label{eq:factor f}
\end{align}
\end{subequations}
\paragraph{The Runkel-Watts string.}
With the two CFTs in hand, we can construct the Runkel-Watts string (RWS) by tensoring them together with the usual $\mathfrak{b}\mathfrak{c}$-ghost system to define a critical bosonic string theory. The worldsheet CFT is
\begin{equation}
\begin{array}{c}
\text{Runkel-Watts CFT} \\ \text{$26 - c\leq 1$}
\end{array}
\ \otimes\  
\begin{array}{c} \text{Liouville CFT} \\ \text{$c \geq 25$} \end{array} \ \otimes\  
\begin{array}{c} \text{$\mathfrak{b}\mathfrak{c}$-ghosts} \\ \text{$c= -26$} \end{array}\,.
\label{eq:RW string}
\end{equation}
where the Liouville central charge is $c=1+6(b+b^{-1})^2$. Vanishing of the total central charge imposes
\begin{equation}
b=\beta=\sqrt{\frac{q'}{q}} \,.
\end{equation}
Thus, the Runkel-Watts string is a family of critical string theories labeled by pairs of coprime positive integers $(q,q')$.

The physical spectrum of the Runkel-Watts string consists of states obtained by combining Runkel-Watts primaries with Liouville primaries, subject to the on-shell condition that the total conformal weight is one. This condition sets the Liouville momentum equal to the Runkel-Watts momentum, $p=P$, and hence vertex operators representing on-shell closed string states take the form
\begin{equation}
\mathcal{V}_p = \mathcal{N}_p \, \widehat{V}_p V_p \,, \qquad p \in \RR_{\geq 0} \,,
\end{equation}
where the factor $\mathcal{N}_p$ is a normalization constant that we will fix later.
Note that in order to define these vertex operators, we have to analytically continue the Liouville momenta to $p\in \RR_{\geq 0}$, which is outside the usual physical spectrum of Liouville theory. 

The observables in the Runkel-Watts string are computed by worldsheet diagrams as usual in string perturbation theory. For a worldsheet of genus $g$ with $n$ external punctures, we define
\begin{align} \label{eq:RW string amplitudes def}
\mathsf{RW}_{g,n}(p_1,\ldots,p_n)
= \biggr( \prod_{j=1}^n \mathcal{N}_{p_j} \biggr)
\int_{\mathcal{M}_{g,n}} Z_{\mathrm{gh}}\,
\langle V_{p_1}\cdots V_{p_n} \rangle_g
\langle \widehat{V}_{p_1}\cdots \widehat{V}_{p_n} \rangle_g \,,
\end{align}
where $\langle V_{p_1}\cdots V_{p_n} \rangle_g$ is the correlation function at genus-$g$ in Liouville CFT and $\langle \widehat{V}_{p_1}\cdots \widehat{V}_{p_n} \rangle_g$ is the corresponding Runkel-Watts CFT correlator. The factor $Z_{\mathrm{gh}}$ is the correlator of the $\mathfrak{b}\mathfrak{c}$-ghost system, and the product of worldsheet CFT correlators is integrated over $\mathcal{M}_{g,n}$, the moduli space of genus-$g$ Riemann surfaces with $n$ punctures.

The string amplitudes \eqref{eq:RW string amplitudes def} can be in principle computed by a direct moduli integration of the worldsheet CFT correlators, upon decomposing the latter into conformal blocks and using the known CFT data. However, as we will see in the next section, we will exploit the relation between the RW-CFT and Liouville CFT of the previous paragraph in order to obtain the RW string amplitudes by a suitable limit of the amplitudes of the complex Liouville string. 

\paragraph{The sphere three-point string amplitude from $\mathbb{C}$LS.}
The first nontrivial string amplitude of the Runkel-Watts string \eqref{eq:RW string} is the sphere three-point amplitude. Since the worldsheet diagram contains no nontrivial moduli, this amplitude is simply the product of the three-point functions of the two constituent CFTs, up to the normalization factors $\mathcal{N}_{p_j}$ and the sphere normalization constant $C_{S^2}$,
\begin{align}
\mathsf{RW}_{0,3}(p_1,p_2,p_3) &= C_{S^2} \biggr( \prod_{j=1}^3 \mathcal{N}_{p_j} \biggr) C_b(p_1,p_2,p_3) \big( C_b(p_1,p_2,p_3) \big)^{-1} \sigma(p_1,p_2,p_3) \nonumber\\
&= C_{S^2} \biggr( \prod_{j=1}^3 \mathcal{N}_{p_j} \biggr) \sigma(p_1,p_2,p_3) \,.
\end{align}
and hence is proportional to the non-analytic step function $\sigma(p_1,p_2,p_3)$. 

Let us now describe how this amplitude is recovered from the corresponding $\mathbb{C}$LS amplitude. The latter is given by \cite{Collier:2024kwt}
\begin{align}
\mathsf{A}^{(b)}_{0,3}(p_1,p_2,p_3) &= C^{\mathbb{C}\text{LS}}_{S^2} \biggr( \prod_{j=1}^3 \mathcal{N}^{\mathbb{C}\text{LS}}_{p_j} \biggr) C_b(p_1,p_2,p_3) C_{ib}(ip_1,ip_2,ip_3) \label{eq:CLS 3pt amp 1} \\
&=  \sum_{m=1}^{\infty} \frac{2b(-1)^m \sin(2\pi m bp_1)\sin(2\pi m bp_2)\sin(2\pi m bp_3)}{\sin(\pi m b^2)} \,. \label{eq:CLS 3pt amp 2}
\end{align}
As seen in \eqref{eq:Liouville to RW limit}, to recover the RW string amplitude we need to take the regulated limit of the $\mathbb{C}$LS amplitude as $b$ approaches $\sqrt{q'/q}$ from the upper half-plane.
Let us first compute this limit of the infinite sum representation of the $\mathbb{C}$LS amplitude in \eqref{eq:CLS 3pt amp 2}. 
First, note that the denominator $\sin(\pi m b^2)$ has a simple zero at $b^2=q'/q$ only if $m$ is a multiple of $q$, so only such subsequence of terms in the sum contributes to the residue. Let's set $m=q k$, and sum $\sum_{k=1}^{\infty}$. Defining the shorthand $x_j\equiv \sqrt{qq'}p_j$, we then have
\begin{align} \label{eq:CLS 3pt limit 1}
\lim_{\varepsilon\to0} i\varepsilon q\,\mathsf{A}^{(b)}_{0,3}(p_1,p_2,p_3)& \Big|_{b=\sqrt{\frac{q'}{q}}+i\varepsilon} = \sum_{k=1}^\infty \frac{(-1)^{k(q'+q)}}{\pi k} \prod_{j=1}^3 \sin(2\pi k x_j) \nonumber\\
&=-\frac{1}{4\pi}\sum_{\sigma_2,\sigma_3=\pm} \sigma_2 \sigma_3 \sum_{k=1}^\infty
\frac{\sin\big(2\pi k(x_1 + \sigma_2 x_2 + \sigma_3 x_3 + \tfrac{q'+q}{2})\big)}{k} \nonumber\\
&=\frac{1}{4}\sum_{\sigma_2,\sigma_3=\pm} \sigma_2 \sigma_3 B_1(\{ x_1 + \sigma_2 x_2 + \sigma_3 x_3 + \tfrac{q'+q}{2} \}) \nonumber\\
&=-\frac{1}{4}\sum_{\sigma_2,\sigma_3=\pm} \sigma_2 \sigma_3 \lfloor x_1 + \sigma_2 x_2 + \sigma_3 x_3 + \tfrac{q'+q}{2} \rfloor \,,
\end{align}
where in the third equality we made use of the Bernoulli Fourier series
\begin{align}
    \sum_{k=1}^\infty \frac{\sin(2\pi ky)}{k} = -\pi B_1(\{y\}) \,,
\end{align}
and in the last equality we used that the constant and linear terms in $B_1(\{y\})=y-\lfloor y \rfloor - \tfrac{1}{2}$ drop out after summing over $\sigma_2\sigma_3$. 
Lastly, it can be shown that the remaining alternating floor sum is proportional to the step function $\sigma(p_1,p_2,p_3)$, by the elementary identity proved in appendix~\ref{app:details-of-calculations}.
In fact, we obtain
\begin{align} \label{eq:CLS 3pt limit 2}
    \lim_{\varepsilon\to0} i\varepsilon q\,\mathsf{A}^{(b)}_{0,3}(p_1,p_2,p_3)\Big|_{b=\sqrt{\frac{q'}{q}}+i\varepsilon} = \frac{1}{4} \left( \prod_{j=1}^3 \sgn \sin\big( 2\pi (\sqrt{qq'}p_j + \tfrac{q'+q}{2})\big) \right) \sigma(p_1,p_2,p_3) \,.
\end{align}
Note that we may also take the regularized limit of the RHS of the first equality in \eqref{eq:CLS 3pt amp 1}, applying the limit \eqref{eq:Liouville to RW limit} to the second factor $C_{ib}(ip_1,ip_2,ip_3)$ coming from the second Liouville CFT in the $\mathbb{C}$LS. The first factor coming from the first Liouville CFT is regular in the limit. 
Equating the result to \eqref{eq:CLS 3pt limit 2} gives a simple way to read off the factors $f(P_j)$ in \eqref{eq:Liouville to RW limit}: 
\begin{align}
     C^{\mathbb{C}\text{LS}}_{S^2} A \prod_{j=1}^3 \mathcal{N}^{\mathbb{C}\text{LS}}_{p_j} f(p_j) 
    = \frac{1}{4} \prod_{j=1}^3 \sgn \sin\big( 2\pi (\sqrt{qq'}p_j + \tfrac{q'+q}{2})\big) \,.
\end{align}
Plugging in the values from \cite{Collier:2024kwt},
\begin{align}
    C^{\mathbb{C}\text{LS}}_{S^2} = 32\pi^4 \left( \frac{\sin(\pi b^2)\sin(\pi b^{-2})}{b^{-2}-b^2} \right)^2 \,, 
    \quad 
    \mathcal{N}^{\mathbb{C}\text{LS}}_{p} = \frac{b^{-2}-b^2}{8\sqrt{2}\pi \sin(\pi b^2)\sin(\pi b^{-2})} \frac{\rho_b(p)}{p} \,, \label{eq:CLS CS2 Np}
\end{align}
we obtain the explicit form of the factor $f(P)$ and the constant $A$ given in \eqref{eq:factor f}.

In the rest of the paper, we will fix the normalization constants $\mathcal{N}_p$ and $C_{\Sigma_g}$ of the Runkel-Watts string in such a way that the relation between the RW string amplitudes and the $\mathbb{C}$LS amplitudes is as simple as possible. The only remaining overall factor will be a universal power of $q$, which we included on the LHS of \eqref{eq:CLS 3pt limit 1}.
From the case of the three-point amplitude, we see that the string vertex operator normalization $\mathcal{N}_p$ is proportional to $\sgn \sin\big( 2\pi (\sqrt{qq'}p + \tfrac{q'+q}{2})\big)$. More precisely we set
\begin{align}\label{eq:RW calN}
    C_{S^2} \prod_{j=1}^3 \mathcal{N}_{p_j} = \frac{1}{4} \prod_{j=1}^3 \sgn \sin\big( 2\pi (\sqrt{qq'}p_j + \tfrac{q'+q}{2})\big) \,.
\end{align}
In summary, the Runkel-Watts string three-point sphere amplitude is given by
\begin{equation} \label{eq:RW 3pt amplitude}
    \mathsf{RW}_{0,3}(p_1,p_2,p_3) = \frac{1}{4} \left( \prod_{j=1}^3 \sgn \sin\big( 2\pi (\sqrt{qq'}p_j + \tfrac{q'+q}{2})\big) \right) \sigma(p_1,p_2,p_3) \,.
\end{equation}
The additional factor $q$ in the limit \eqref{eq:CLS 3pt limit 2} is the first instance of the factor $q^{2g-2+n}$ which we absorb into the definition of the Runkel-Watts amplitudes, see \eqref{eq:RW amplitude limit}.

\paragraph{Analytic continuation: Definition for small momenta.}
In the worldsheet perturbation theory definition \eqref{eq:RW string amplitudes def}, the Runkel-Watts operators $\widehat V_{P_i}$ are only defined for $P_i\in\RR_{\geq 0}$. The corresponding contour for intermediate Runkel-Watts states is therefore the positive half of the real line. As discussed above, the poles of the inverse DOZZ structure constant on this contour are removed by the chamber factor $\sigma(P_1,P_2,P_3)$, so they do not lead to divergences in the real-momentum integral.

By contrast, the Liouville operators $V_{p_i}$ that appear in the Runkel-Watts string should be viewed as analytic continuations from the original Liouville spectrum $p\in i\RR_{\geq 0}$. The contour $\mathcal C$ for intermediate Liouville momenta is therefore more subtle and depends on the external momenta. Similar questions arise in \CLS{} \cite{Collier:2024kwt} and also ADE minimal strings \cite{Rodriguez:2025rte}. When the external Liouville momenta are purely imaginary, the internal Liouville contour is also purely imaginary, $\mathcal C=i \mathbb R_{\geq 0}$. After analytic continuation of the external momenta, however, poles of the DOZZ structure constants may cross this contour. In that case, the contour prescription is modified by subtracting the contributions of the crossed poles, in order to preserve crossing symmetry of the Liouville correlator. Alternatively, one may smoothly deform the integration contour.

A priori, the crossed poles can introduce analytic features into the correlator. For example, in the $\CC$LS, such crossed poles lead to discontinuities in the amplitudes \cite{Collier:2024kwt}. For the VMS, it was demonstrated in \cite{Khromov:2025awh} for $\mathsf{V}_{0,4}^{(b)}$ that there are in fact no further discontinuities from the pole crossing and $\mathsf{V}_{0,4}^{(b)}$ is an entire function.

For the RW string, the analysis is essentially identical since the structure constants of the Runkel-Watts CFT differ from those of timelike Liouville theory only by the chamber function. This means that the correlation function may be similarly analytically continued outside the original region of convergence. This does not introduce new singularities beyond those that are already present in the structure constants due to the chamber function. 

\begin{figure}[ht]
    \centering
    \begin{subfigure}[b]{.45\textwidth}
        \includegraphics[width=\textwidth]{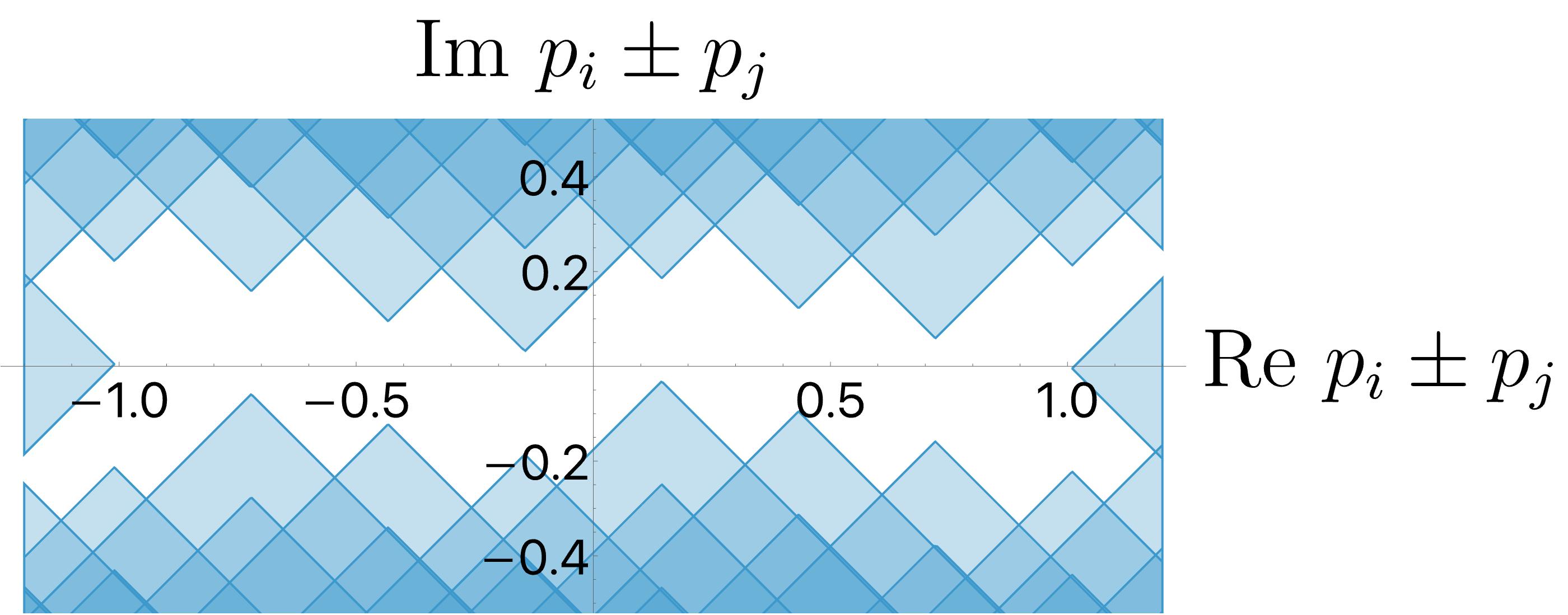}
        \caption{$b = \sqrt{\tfrac{3}{4}}\mathrm{e}^{\frac{\pi i}{100}}$}
    \end{subfigure}
    ~
    \begin{subfigure}[b]{.45\textwidth}
        \includegraphics[width=\textwidth]{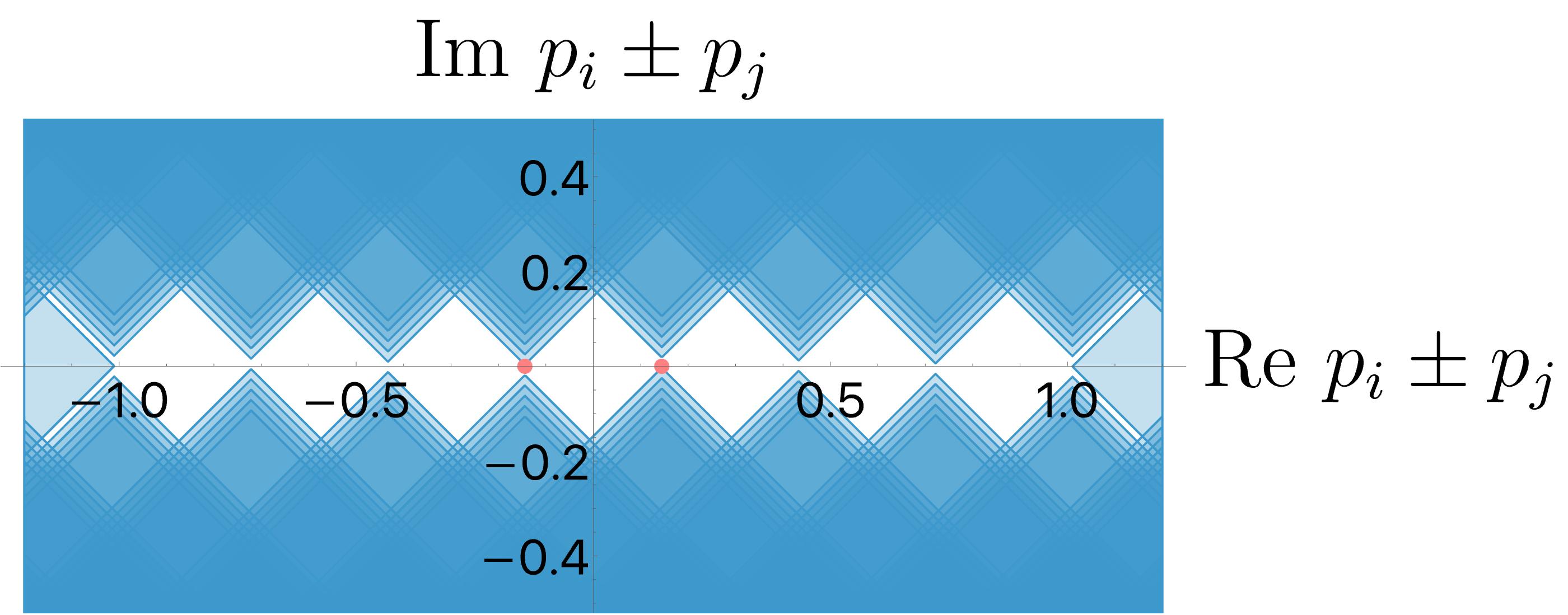}
        \caption{$b = \sqrt{\tfrac{3}{4}}\mathrm{e}^{\frac{\pi i}{1000}}$}
    \end{subfigure} 
    \\
    \begin{subfigure}[b]{.45\textwidth}
        \includegraphics[width=\textwidth]{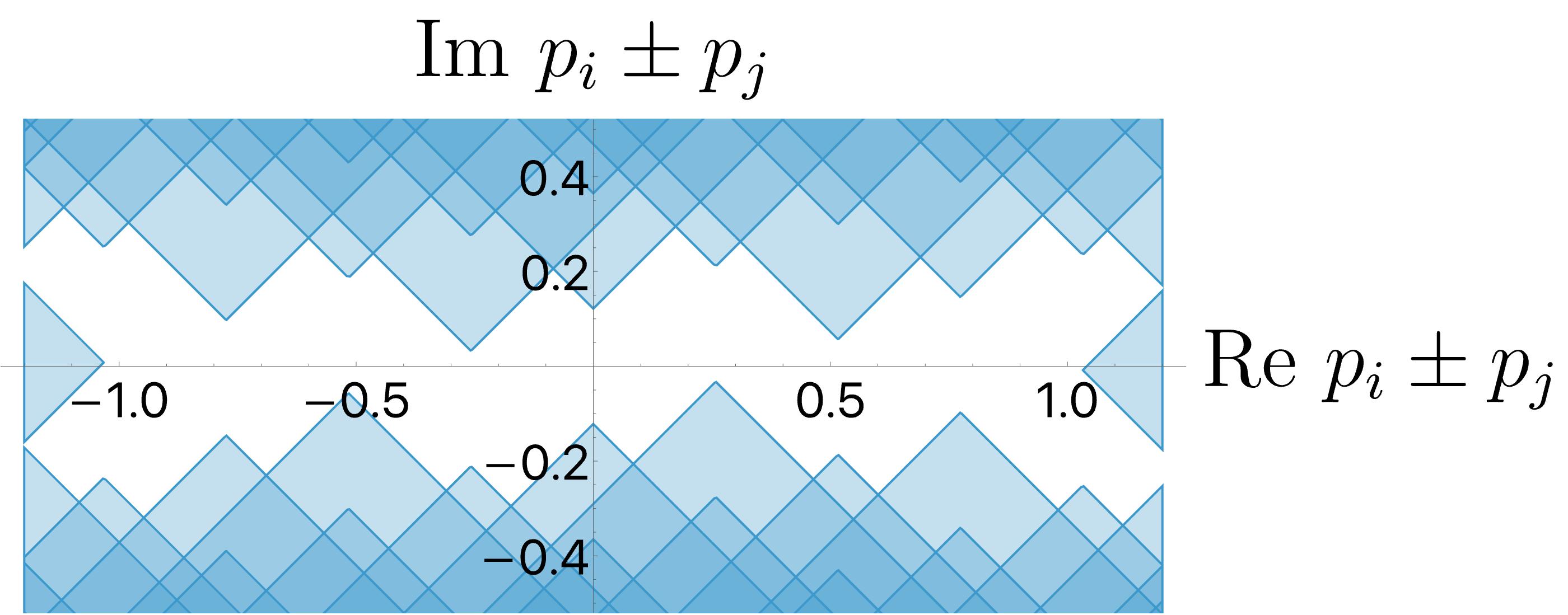}
        \caption{$b = \sqrt{\tfrac{3}{5}}\mathrm{e}^{\frac{\pi i}{100}}$}
    \end{subfigure}
    ~
    \begin{subfigure}[b]{.45\textwidth}
        \includegraphics[width=\textwidth]{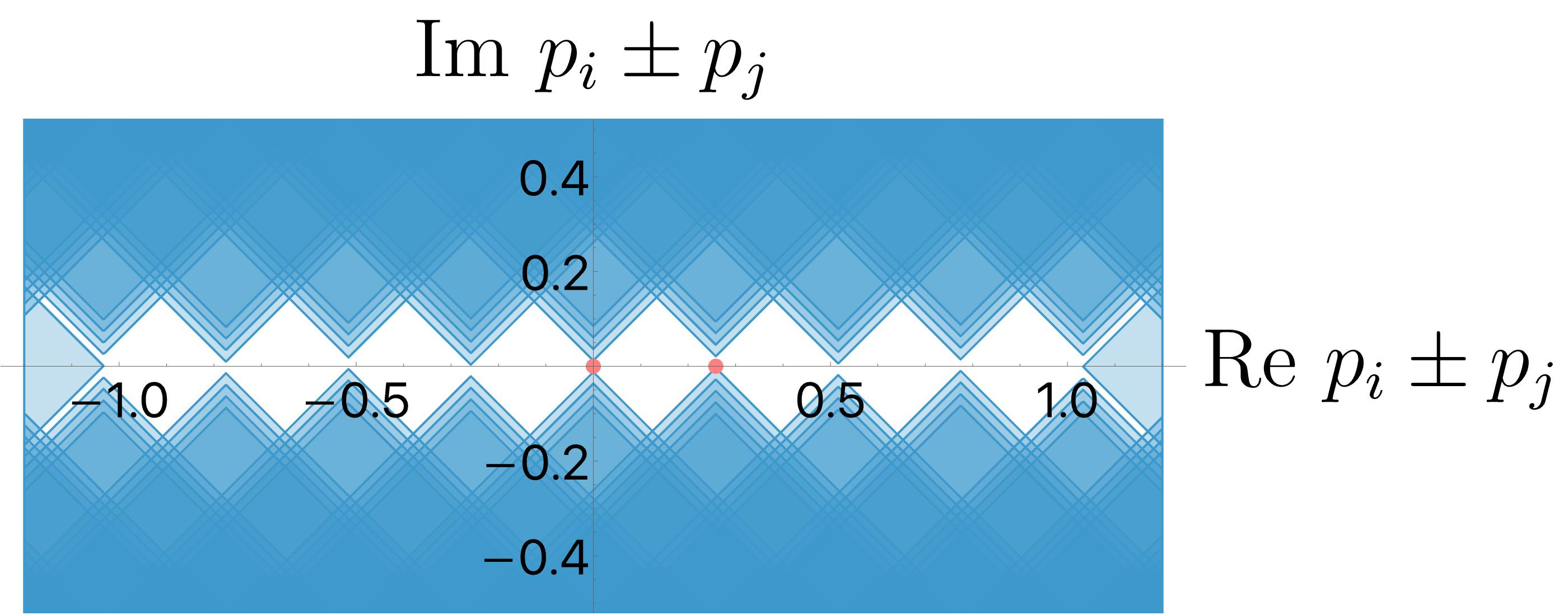}
        \caption{$b = \sqrt{\tfrac{3}{5}}\mathrm{e}^{\frac{\pi i}{1000}}$}
    \end{subfigure}
    \caption{The unshaded area represents the region in the external Liouville momenta where the moduli space integral that defines the $\CC$LS sphere four-point amplitude converges. 
    The shaded regions correspond to 90-degree wedges of divergence emanating from the branch points of the amplitude.
    In the limit that $b\to \sqrt{q'/q}$, the region of convergence pinches off into $q'+q$ rhombi of convergence, lying within the range $-\frac{b+b^{-1}}{2}\leq \Re p_i\pm p_j \leq \frac{b+b^{-1}}{2}$. 
    Infinitely many branch points collide at $p_i\pm p_j = \pm \frac{1}{2\sqrt{qq'}}\mathbb{Z}$ for $q'+q$ odd (sample pink dots in top figures), and at $p_i\pm p_j = 0, \pm \frac{1}{\sqrt{qq'}}\mathbb{Z}$ for $q'+q$ even (sample pink dots in bottom figures).}\label{fig:region of convergence}
\end{figure}

One can also study the analyticity of RWS amplitudes from the \CLS{} perspective; see figure \ref{fig:region of convergence}.
The unshaded region indicates the domain in which the limit defining the Runkel-Watts amplitudes can be taken directly from the \CLS{} amplitudes. Outside this region, the amplitudes are defined by analytic continuation across the corresponding non-analytic loci inherited from the \CLS{} description. 

\subsection{Amplitudes from geometry}\label{subsec:amplitudes-from-geometry}

\paragraph{Intersection number formula for $\mathbb{C}$LS amplitudes.}
We use the intersection-theoretic representation of the $\mathbb{C}$LS amplitudes as the starting point for the general correlators of the Runkel-Watts string. The $\mathbb{C}$LS amplitude can be written as a sum over colored stable graphs \cite[eq. (B.19)]{Collier:2024lys},
\begin{align}
    \mathsf{A}_{g,n}^{(b)}(\boldsymbol{p})&= \sum_{\Gamma \in \mathcal{G}_{g,n}^\infty}\frac{1}{|\text{Aut}(\Gamma)|}\prod_{v \in \mathcal{V}_\Gamma} \left(\frac{b (-1)^{m_v}}{\sqrt{2}\sin(\pi m_v b^2)}\right)^{2g_v-2+n_v} \!\! \int_{\bM_\Gamma} \prod_{v \in \mathcal{V}_\Gamma}\ex^{\frac{b^2+b^{-2}}{4} \kappa_1-\sum_k\frac{B_{2k}\kappa_{2k}}{(2k)(2k)!}} \nonumber\\
    &\quad\times\!\prod_{(\bullet,\circ) \in \mathcal{E}_\Gamma} \sum_{d=0}^\infty \frac{\Gamma(d+\tfrac{3}{2})}{\sqrt{\pi}(\pi b)^{2d+2}}\left(\frac{\delta_{m_\bullet\ne m_\circ}}{(m_\bullet-m_\circ)^{2d+2}}-\frac{1}{(m_\bullet+m_\circ)^{2d+2}}\right)(\psi_\bullet+\psi_\circ)^d \nonumber\\
    &\quad\times \prod_{i=1}^n \ex^{-p_i^2 \psi_i} \sqrt{2}\sin(2\pi m_i b p_i)\,.\label{eq:CLS intersection expression}
\end{align}
Here $\boldsymbol{p}=(p_1,\ldots,p_n)$, and $\mathcal{G}_{g,n}^\infty$ denotes the set of connected stable graphs with $n$ external legs and vertex colors $m_v\in \ZZ_{\geq 1}$. Each vertex $v$ carries a genus $g_v$ and valence $n_v$, including both internal and external half-edges, with total genus
\be
    g=\sum_{v\in \mathcal{V}_\Gamma}g_v+\dim H^1(\Gamma,\RR)\,,
\ee
where the second contribution is the loop number of the graph. 
The stability condition is the usual one: $n_v\geq 3$ for genus-zero vertices and $n_v\geq 1$ for genus-one vertices. We write $\bM_\Gamma=\prod_{v\in \mathcal{V}_\Gamma}\bM_{g_v,n_v}$, and the automorphism factor is that of the underlying graph. Passing to the compactification is essential, since the formula is an intersection number built from the standard $\psi$- and $\kappa$-classes on the Deligne-Mumford moduli spaces.

The three lines of \eqref{eq:CLS intersection expression} have the structure of Feynman rules. The first line assigns to each vertex a color-dependent weight together with the exponential of the $\kappa$-classes. The second line is the edge factor: for an edge joining the half-edges $\bullet$ and $\circ$, the adjacent colors are $m_\bullet$ and $m_\circ$, while $\psi_\bullet$ and $\psi_\circ$ denote the cotangent line classes at the two nodal markings. The final line contains the external leg factors; $m_i$ is the color of the vertex on which the $i$th external leg ends, and $\psi_i$ is the corresponding external cotangent class. In evaluating the integral, only the component of total cohomological degree $2\dim_\CC\bM_\Gamma$ contributes.

The formula was presented in \cite{Collier:2024lys}, where it was shown to obey highly nontrivial properties that were derived from the worldsheet perspective in \cite{Collier:2024kwt}. It also agrees with the explicit analytic bootstrap of the low-genus amplitudes in \cite{Collier:2024kwt}, together with direct numerical checks of these amplitudes. Further compelling evidence comes from \cite{Collier:2026pxi}, where $c=1$ string amplitudes were extracted from the same $\mathbb{C}$LS formula and shown to match the dual matrix quantum mechanics. Here we will instead take the rational Runkel-Watts limit of \eqref{eq:CLS intersection expression}.

\paragraph{The Runkel-Watts limit.}
We now take the rational limit of \eqref{eq:CLS intersection expression}. Let
\be
    b_0=\sqrt{\frac{q'}{q}}\,,
\ee
with $q$ and $q'$ coprime positive integers, and approach the rational point from the upper half-plane as $b=b_0+i\varepsilon$ with $\varepsilon>0$. We define
\be\label{eq:RW amplitude limit}
    \mathsf{RW}_{g,n}(\boldsymbol{p})
    =\lim_{\varepsilon\to 0}(iq\varepsilon)^{2g-2+n}
    \mathsf{A}_{g,n}^{(b_0+i\varepsilon)}(\boldsymbol{p})\,.
\ee
The power of $i\varepsilon$ is dictated by the vertex factors in \eqref{eq:CLS intersection expression}. The additional power $q^{2g-2+n}$ is needed to match to our previous conventions and in particular will make the expression $q \leftrightarrow q'$ symmetric. It corresponds to a renormalization of the string coupling. Since $q$ and $q'$ are coprime, $\sin(\pi m b_0^2)=0$ if and only if $m\in q\ZZ$. Thus the leading term in \eqref{eq:RW amplitude limit} receives contributions only from graphs for which every vertex color is divisible by $q$. Writing the original $\mathbb{C}$LS color as $q m_v$ and reusing $m_v$ for the reduced color, one finds
\be
    \lim_{\varepsilon\to 0}i\varepsilon q
    \frac{(b_0+i\varepsilon)(-1)^{q m_v}}{\sqrt{2}\sin(\pi q m_v (b_0+i\varepsilon)^2)}
    =
    \frac{(-1)^{(q+q')m_v}}{2\sqrt{2}\pi\,m_v}\,.
\ee
All remaining factors in \eqref{eq:CLS intersection expression} are regular at $b=b_0$ and can be evaluated there. The edge denominators acquire the additional powers of $q$ from the same relabeling of the colors, which combine with $b_0$ to give $\sqrt{qq'}$.

We therefore obtain the following intersection number expression:
\begin{align} \label{eq:RW intersection number formula}
\mathsf{RW}&_{g,n}(\boldsymbol{p}) 
\nonumber\\
&=\sum_{\Gamma\in \mathcal{G}_{g,n}^\infty} \frac{1}{|\Aut(\Gamma)|} \prod_{v\in \mathcal{V}_\Gamma} 
\left( \frac{(-1)^{(q+q')m_v}}{2\sqrt{2}\pi m_v} \right)^{2g_v-2+n_v}
\!\!\int_{\bM_\Gamma}
\prod_{v\in \mathcal{V}_\Gamma}
\mathrm{e}^{
\frac{1}{4}\left(\frac{q'}{q}+\frac{q}{q'}\right)\kappa_1 - \sum_{k\geq 1}\frac{B_{2k}\kappa_{2k}}{(2k)(2k)!}
}
\nonumber\\
&\quad\times
\prod_{(\bullet,\circ)\in \mathcal{E}_\Gamma}
\sum_{d=0}^\infty
\frac{\Gamma(d+\frac{3}{2})}{\sqrt{\pi}\,(\pi\sqrt{qq'})^{2d+2}}
\left(
\frac{\delta_{m_\bullet\ne m_\circ}}{(m_\bullet-m_\circ)^{2d+2}}
-\frac{1}{(m_\bullet+m_\circ)^{2d+2}}
\right)
(\psi_\bullet+\psi_\circ)^d
\nonumber\\
&\quad\times
\prod_{i=1}^n
\ex^{-p_i^2\psi_i} \sqrt{2}\sin(2\pi \sqrt{qq'}\,m_i p_i)\,.
\end{align}

\paragraph{From intersection number expression to topological recursion.}
The stable graph formula \eqref{eq:RW intersection number formula} is also the natural starting point for the topological recursion description. The basic point is that topological recursion reconstructs the resolvent differentials by summing residues at the branch points of the map $\xx$. On the other hand, the intersection-theory theorem of \cite[Theorem 4.1]{Eynard:2011ga} expresses these differentials as a sum over colored stable graphs, with a closely related formulation in terms of the Givental formalism given in \cite{Dunin-Barkowski:2012kbi}. The colors are precisely the branch-point labels, while the edge and leg factors are fixed by the local recursion data. In the present case the computation is essentially the same as for $\CC$LS, whose derivation is explained in detail in \cite[App. B]{Collier:2024lys}.

We will therefore not repeat the reconstruction in detail. Comparing \eqref{eq:RW intersection number formula} with the universal stable graph expansion determines the topological recursion data, with the integer color $m$ labeling the corresponding branch point. This leads to the Runkel-Watts spectral curve \eqref{eq:RW spectral curve intro} that we describe in section \ref{subsec:dual-spectral-curve}.

\paragraph{Trading sums for integrals: momentum space Feynman rules.} One can make more analytic structure of \eqref{eq:RW intersection number formula} manifest by the following manipulations. We have the basic identity
\be 
\sum_{m \ne 0} \frac{\mathrm{e}^{2\pi i m x}}{m^d}=-\frac{(2\pi i)^d}{d!}\, B_d(\{x\})\ , \label{eq:Bernoulli polynomial sum identity}
\ee
where $B_d$ denotes the Bernoulli polynomial. We will apply this identity twice, both for the edges and the vertices. For every half edge, we first introduce a new sign $\sigma$ to account for the two terms in \eqref{eq:RW intersection number formula}. We can then write
\begin{align} 
&\frac{\delta_{m_\bullet\ne m_\circ}}{(m_\bullet-m_\circ)^{2d+2}}
-\frac{1}{(m_\bullet+m_\circ)^{2d+2}}\nonumber\\
&\qquad=-\frac{1}{2}\sum_{\sigma_\bullet,\sigma_\circ=\pm 1} \sigma_\bullet \sigma_\circ \frac{\delta_{\sigma_\bullet m_\bullet +\sigma_\circ m_\circ\ne 0}}{(\sigma_\bullet m_\bullet+\sigma_\circ m_\circ)^{2d+2}}\\
&\qquad=\frac{(2\pi i)^{2d+2}\sqrt{qq'}}{2(2d+2)!} \sum_{\sigma_\bullet,\sigma_\circ =\pm 1} \sigma_\bullet \sigma_\circ \int_{\RR\big/\frac{1}{\sqrt{qq'}}\ZZ} \d k\, B_{2d+2}(\{k \sqrt{qq'}\}) \, \mathrm{e}^{2\pi i k \sqrt{qq'}(\sigma_\bullet m_\bullet+\sigma_\circ m_\circ)}\ .
\end{align}
The last step line can be confirmed by plugging in the Fourier expansion \eqref{eq:Bernoulli polynomial sum identity} and noticing that the $k$-integral projects to one Fourier mode.
This factorizes the $\sigma_\bullet$ and $\sigma_\circ$ dependence to be concentrated in the vertices. We can also rewrite the sine functions of the external legs in \eqref{eq:RW intersection number formula} as a sum over two exponential functions. This sum is also parametrized by an assignment of $\sigma_i$. Overall, we hence naturally obtain a sum over $\boldsymbol{\sigma} \in \mathcal{H}_\Gamma$, a sign associated to every half edge, both internal and external, together with an integral over momenta associated to every edge. For a half edge $h$, we denote $p_h=k_e$ if $h$ is part of the internal edge $e$ and $p_h=p_i$ if $h$ is the external half edge with label $i$. With this, \eqref{eq:RW intersection number formula} becomes
\begin{align} \label{eq:RW intersection number formula 2}
\mathsf{RW}_{g,n}(\boldsymbol{p}) 
&=\sum_{\Gamma\in \mathcal{G}_{g,n}^\infty} \frac{(\frac{1}{\sqrt{2}i})^n}{|\Aut(\Gamma)|} \sum_{\boldsymbol{\sigma} \in \mathcal{H}_\Gamma}   \int_{\RR\big/ \frac{1}{\sqrt{qq'}} \ZZ} \d^{|\mathcal{E}_\Gamma|} \boldsymbol{k}\prod_{v\in \mathcal{V}_\Gamma} 
\left( \frac{1}{2\sqrt{2}\pi m_v} \right)^{2g_v-2+n_v} \prod_{h \in \mathcal{H}_v} \sigma_h
\nonumber\\
&\quad\times \int_{\bM_\Gamma}
\prod_{v\in \mathcal{V}_\Gamma}
\mathrm{e}^{
\frac{1}{4}\left(\frac{q'}{q}+\frac{q}{q'}\right)\kappa_1 - \sum_{k\geq 1}\frac{B_{2k}\kappa_{2k}}{(2k)(2k)!}
} \mathrm{e}^{2\pi i m_v  (\sqrt{qq'}\sum_{h \in \mathcal{H}_v}\sigma_h p_h+\frac{n_v}{2}(q+q'))}
\nonumber\\
&\quad\times
\prod_{(\bullet,\circ)\in \mathcal{E}_\Gamma}
\sum_{d=0}^\infty
\frac{(-1)^{d+1}}{2 (d+1)!\,(qq')^{d+\frac{1}{2}}}\, B_{2d+2}(\{\sqrt{qq'} k_e\})
(\psi_\bullet+\psi_\circ)^d \prod_{i=1}^n
\ex^{-p_i^2\psi_i} \ .
\end{align}
One can now perform the sum over the vertex colors by using \eqref{eq:Bernoulli polynomial sum identity} again. For this, we notice that the sum over $m_v$ is symmetric under $m_v \to -m_v$, since this sign flip can be compensated by also flipping the signs of all $\sigma_h$ for $h \in \mathcal{H}_v$, the half edges emanating from $v$. We can thus extend the sum over $m_v$ to $\ZZ \setminus \{0\}$ and compensate by a factor of $\frac{1}{2}$. This reduces the sum over colored stable graphs $\mathcal{G}_{g,n}^\infty$ to a sum over ordinary stable graphs $\mathcal{G}_{g,n}$. Using that $\sum_{v}(2g_v-2+n_v)=2g-2+n$, we obtain
\begin{align} \label{eq:RW intersection number formula 3}
\mathsf{RW}_{g,n}(\boldsymbol{p}) 
&=\sum_{\Gamma\in \mathcal{G}_{g,n}} \frac{(-2)^{1-g}2^{-n}}{|\Aut(\Gamma)|} \sum_{\boldsymbol{\sigma} \in \mathcal{H}_\Gamma}  \int_{\RR\big/ \frac{1}{\sqrt{qq'}} \ZZ} \d^{|\mathcal{E}_\Gamma|} \boldsymbol{k}\nonumber\\
&\quad\times \prod_{v\in \mathcal{V}_\Gamma}
\frac{-\prod_{h \in \mathcal{H}_v}  \sigma_h }{2(2g_v-2+n_v)!}\, B_{2g_v-2+n_v}\Big(\!\Big\{\sqrt{qq'}\sum_{h \in \mathcal{H}_v} \sigma_h p_h+\frac{n_v}{2}(q+q')\Big\}\!\Big) \nonumber\\
&\quad\times \int_{\bM_\Gamma}
\prod_{v\in \mathcal{V}_\Gamma}
\mathrm{e}^{
\frac{1}{4}\left(\frac{q'}{q}+\frac{q}{q'}\right)\kappa_1 - \sum_{k\geq 1}\frac{B_{2k}\kappa_{2k}}{(2k)(2k)!}
} \prod_{i=1}^n
\ex^{-p_i^2\psi_i} 
\nonumber\\
&\quad\times
\prod_{(\bullet,\circ)\in \mathcal{E}_\Gamma}
\sum_{d=0}^\infty
\frac{(-1)^{d+1}}{2 (d+1)!\,(qq')^{d+\frac{1}{2}}}\, B_{2d+2}(\{\sqrt{qq'} k_e\})
(\psi_\bullet+\psi_\circ)^d  \ .
\end{align}
We thus traded the infinite sum over colors with a finite sum over signs, together with loop integral associated to all the edges. One can optionally rewrite this further by using the same manipulation as in \cite{Collier:2026pxi}, which expresses
\begin{multline}
\sum_{d=0}^\infty
\frac{(-1)^{d+1}}{2 (d+1)!\,(qq')^{d+\frac{1}{2}}}\, B_{2d+2}(\{\sqrt{qq'} k_e\})
(\psi_\bullet+\psi_\circ)^d\\
=\frac{1}{2}\sum_{\ell_e \in \ZZ} \Big| k_e+\frac{\ell_e}{\sqrt{qq'}}\Big|^{-s}\, \mathrm{e}^{-(\psi_\bullet+\psi_\circ)(k_e+\frac{\ell_e}{\sqrt{qq'}})^2}\bigg|_{s=-1}\ .
\end{multline}
The RHS should be viewed as a power series in the $\psi$-classes. It defines an analytic function in $s$ and after performing the sum over $\ell_e$ and its value is defined by analytic continuation to $s=-1$. 
This rewriting also factorizes the $\psi$-classes associated to the edges and the moduli space integral over each vertex can be performed in terms of the VMS quantum volumes, which are given by \cite{Collier:2023cyw}
\be \label{eq:VMS quantum volume definition RW}
    \mathsf{V}^{(b)}_{g,n}(P_1,\ldots,P_n)
    =
    \int_{\bM_{g,n}}
    \exp\left(
    \frac{b^2+b^{-2}}{4}\kappa_1
    +\sum_{j=1}^n P_j^2\psi_j
    -\sum_{k\geq 1}\frac{B_{2k}\kappa_{2k}}{(2k)(2k)!}
    \right)\, .
\ee
We will use the analytic continuation of \eqref{eq:VMS quantum volume definition RW} to imaginary momenta, which is immediate since the right hand side is a polynomial in the $P_j^2$.

This leads to
\begin{align} \label{eq:RW intersection number formula 4}
\mathsf{RW}_{g,n}(\boldsymbol{p}) 
&=\sum_{\Gamma\in \mathcal{G}_{g,n}} \frac{(-2)^{1-g}2^{-n}}{|\Aut(\Gamma)|} \sum_{\boldsymbol{\sigma} \in \mathcal{H}_\Gamma}  \int_{\RR\big/ \frac{1}{\sqrt{qq'}} \ZZ} \d^{|\mathcal{E}_\Gamma|} \boldsymbol{k}\sum_{\boldsymbol{\ell}\in \ZZ^{|\mathcal{E}_\Gamma|}} \frac{1}{2^{|\mathcal{E}_\Gamma|}}\nonumber\\
&\quad\times \prod_{v\in \mathcal{V}_\Gamma}
\frac{-\prod_{h \in \mathcal{H}_v}  \sigma_h }{2(2g_v-2+n_v)!}\, B_{2g_v-2+n_v}\Big(\!\Big\{\sqrt{qq'}\sum_{h \in \mathcal{H}_v} \sigma_h p_h+\frac{n_v}{2}(q+q')\Big\}\!\Big) \nonumber\\
&\quad\times \prod_{v \in \mathcal{V}_\Gamma} \mathsf{V}_{g_v,n_v}^{(\sqrt{q'/q})} \Big(i \Big(\boldsymbol{p}_v+\frac{\boldsymbol{\ell}_v}{\sqrt{qq'}}\Big)\Big) \prod_{e \in \mathcal{E}_\Gamma} \Big|k_e+\frac{\ell_e}{\sqrt{qq'}}\Big|^{-s} \bigg|_{s=-1}  \ .
\end{align}
The analytic continuation in $s$ has to be performed after the sum over $\ell$, but before the integral over the loop momenta. However, we can combine the sum over $\boldsymbol{\ell}$ and the integral over $\boldsymbol{k}$ into a regularized integral as follows:
\be 
\int'\d p\,  f(p):=\int_0^1 \d p\, f(p)+\int_1^\infty \d p\, |p|^{-s-1} f(p) \Big|_{s=-1}\ ,
\ee
where $f(p)$ is a polynomially growing function in $p$. Here we choose $\Re s$ large enough so that the integral over $p$ converges. We then analytically continue the result to $s=-1$. The split of the integral is introduced so that the regulator $|p|^{-s-1}$ does not introduce a spurious pole at $p=0$. This definition of the regularized integral is equivalent to the one used in \cite{Collier:2024lys}. We can thus rewrite \eqref{eq:RW intersection number formula 4} further as follows,
\begin{align} \label{eq:RW intersection number formula 5}
\mathsf{RW}_{g,n}(\boldsymbol{p}) 
&=\sum_{\Gamma\in \mathcal{G}_{g,n}} \frac{1}{|\Aut(\Gamma)|} \sum_{\boldsymbol{\sigma} \in \mathcal{H}_\Gamma}  \prod_{e \in \mathcal{E}_\Gamma} \int' (-2 |p_e|\d p_e) \prod_{v\in \mathcal{V}_\Gamma}
\frac{(-2)^{-g_v}2^{-n_v}\prod_{h \in \mathcal{H}_v}  \sigma_h }{(2g_v-2+n_v)!}\nonumber\\
&\quad\times B_{2g_v-2+n_v}\Big(\!\Big\{\sqrt{qq'}\sum_{h \in \mathcal{H}_v} \sigma_h p_h+\frac{n_v}{2}(q+q')\Big\}\!\Big)\!  \prod_{v \in \mathcal{V}_\Gamma} \mathsf{V}_{g_v,n_v}^{(\sqrt{q'/q})} (i \boldsymbol{p}_v)  \ .
\end{align}
We used the fact that thanks to the identities $\sum_{v\in\mathcal{V}}(2g_v-2+n_v)=2g-2+n$ and
$\sum_{v\in\mathcal{V}}n_v = n+2|\mathcal{E}|$, one has
\begin{equation}
    (-2)^{1-g}2^{-n} = (-2)^{|\mathcal{E}_\Gamma|}\prod_{v\in\mathcal{V}}(-2)^{1-g_v}2^{-n_v}~.
\end{equation}
\paragraph{The vertex factor and $\mathrm{SU}(2)$ Yang-Mills theory.} Let us comment a bit more about the vertex factor in the second line of \eqref{eq:RW intersection number formula 5}, which takes the form
\begin{align} 
\mathsf{A}_{g,n}^\mathrm{TQFT}(\boldsymbol{\theta})&=(-2)^{1-g}2^{-n}\sum_{\boldsymbol{\sigma} \in \{\pm \}^n} \frac{-\prod_{i=1}^n \sigma_i}{2(2g-2+n)!} B_{2g-2+n} \Big(\Big\{\sum_{i=1}^n \sigma_i \theta_i \Big\}\Big)\nonumber\\
&=2^{3-3g-n}\pi^{2-2g-n}\prod_{i=1}^n  \sin(2\pi \theta_i) \sum_{m=1}^\infty \frac{\prod_{i=1}^n \chi_m^{\mathrm{SU}(2)}(\theta_i)}{m^{2g-2+n}}\ . \label{eq:SU(2) YM partition function}
\end{align}
where
\be 
\theta_h=\sqrt{qq'}\, p_h+\frac{1}{2}(q+q')\ , \label{eq:theta p relation}
\ee
and
\be 
\chi_m^{\mathrm{SU}(2)}(\theta)=\frac{\sin(2\pi m \theta)}{\sin(2\pi \theta)}
\ee
is the $m$-dimensional $\mathrm{SU}(2)$ character where $\theta \in \RR/\ZZ$ plays the role of the fugacity.

The expression \eqref{eq:SU(2) YM partition function} coincides with $\mathrm{SU}(2)$ Yang-Mills theory. The sum over $m$ corresponds to the sum over all irreducible representations of $\mathrm{SU}(2)$ of which there is precisely one of every dimension $m$. The prefactors are normalizations that can be absorbed into the vertex operator normalizations as well as the Euler counterterm. Thus we can think of the Runkel-Watts string as $\mathrm{SU}(2)$ Yang-Mills theory coupled to gravity. $q$ and $q'$ only enter in the specific way the theory is coupled to gravity. For the complex Liouville string, the underlying TQFT is $q$-deformed $\mathrm{SU}(2)_q$ Yang-Mills theory \cite{Collier:2024lys} and thus the limit $b \to \sqrt{\frac{q'}{q}}$ somewhat surprisingly takes the semiclassical limit on the TQFT independently of $q$ and $q'$.

\paragraph{Summary.} With this identification, we deduce the conceptually satisfying formula 
\begin{multline}
    \mathsf{RW}_{g,n}(\boldsymbol{p})=\sum_{\Gamma \in \mathcal{G}_{g,n}} \frac{1}{|\text{Aut}(\Gamma)|} \prod_{e \in \mathcal{E}_\Gamma} \int'(-2 |k_e|\d k_e) \\
    \times \prod_{v \in \mathcal{V}_\Gamma} \mathsf{A}_{g_v,n_v}^\text{TQFT}\big(\sqrt{qq'}\boldsymbol{p}_v+\tfrac{1}{2}(q+q')\big) \mathsf{V}_{g_v,n_v}^{(\sqrt{q'/q})}(i \boldsymbol{p}_v)\,.\label{eq:RW intersection number formula 6}
\end{multline}
In particular, in this form the explicit $\psi$- and $\kappa$-class expression has disappeared; all moduli-space dependence is packaged into VMS quantum volumes. Thus, as in the $\CC$LS formula \cite[App. B]{Collier:2024lys} and the $c=1$ string formula \cite{Collier:2026pxi} and the $(2,p)$ minimal string formula \cite{Artemev:2025pvk}, the VMS volumes appear as the universal vertices of the Feynman rules. This gives another sense in which the VMS provides the common skeleton for these low-dimensional string theories, now including the Runkel-Watts string.

\paragraph{$q=q'=1$ case.} The special case for $q=q'=1$ is particularly interesting as it corresponds to the case where no linear dilaton potential is turned on along the Runkel-Watts theory. We can thus view it as a deformed background of the $c=1$ string. The intersection number expression \eqref{eq:RW intersection number formula 6} also becomes particularly simple in this case and simply reads
\be 
    \mathsf{RW}_{g,n}(\boldsymbol{p})=\sum_{\Gamma \in \mathcal{G}_{g,n}} \frac{1}{|\text{Aut}(\Gamma)|} \prod_{e \in \mathcal{E}_\Gamma} \int'(-2 |k_e|\d k_e) \prod_{v \in \mathcal{V}_\Gamma} \mathsf{A}_{g_v,n_v}^\text{TQFT}(\boldsymbol{p}_v)\mathsf{V}_{g_v,n_v}^{(1)}(i \boldsymbol{p}_v)\,,\label{eq:RW intersection number formula q=qp=1}
\ee
i.e.\ the external Liouville momenta get directly identified with the $\mathrm{SU}(2)$ monodromies.
This is identical to the formula for the $c=1$ string discussed in \cite{Collier:2026pxi}, except that the TQFT is $\mathrm{SU}(2)$ Yang-Mills theory instead of $\mathrm{U}(1)$ Yang-Mills theory (for which $\mathsf{A}_{g,n}^\text{TQFT}(\boldsymbol{p})=\delta_{\sum_i p_i \in \ZZ}$).

This similarity with the $c=1$ string suggests that the Runkel-Watts string may admit an interesting target-space interpretation. 
A possible hint in this direction comes from \cite{Mazel:2024alu}, which studied the perturbatively marginal deformation of a non-compact Euclidean free boson by $\lambda \cos(\sqrt{2}X)$ using closed string field theory. 
In order to compensate for the change of matter central charge along the deformation, the setup of \cite{Mazel:2024alu} includes an auxiliary linear dilaton sector. 
The resulting string field configuration has non-trivial tachyon and dilaton profiles, with the dilaton diverging at two finite locations in the target space. 
It was proposed in \cite{Mazel:2024alu} that, in the limit $\lambda\to 0$, the CFT describing the neighborhood of each end of space is the $c=1$ Runkel-Watts CFT. 
More generally, away from the strict $\lambda\to 0$ limit, the same construction suggests a finite-interval background bounded by Runkel-Watts walls, with the $A_k$ minimal models appearing at the corresponding large-$k$ values of $\lambda\sim 1/k$.

From the perspective of the present paper, this suggests a possible window into a target-space interpretation for the Runkel-Watts string. 
Replacing the auxiliary linear dilaton sector of \cite{Mazel:2024alu} by a full Liouville CFT sector, one is naturally led to the worldsheet theory studied here. 
It would be interesting to understand whether this relation can be made more precise, and in particular whether the tachyon-dilaton background of \cite{Mazel:2024alu} admits a Lorentzian continuation. 
Such a continuation could provide a target-space interpretation of the amplitudes $\mathsf{RW}_{g,n}$ as asymptotic observables, analogous to an $S$-matrix and similar in spirit to the interpretation of VMS amplitudes considered in \cite{Rodriguez:2023kkl,Rodriguez:2023wun}.

\paragraph{More on $\SU(2)$ Yang-Mills theory.}
An orthogonal route for the relation to $\SU(2)$ Yang-Mills theory and a target space interpretation at least in the case of $q=q'=1$ is provided by \cite{Gaberdiel:2011aa}, which realizes the Runkel-Watts theory at $q=q'=1$ as a continuous orbifold of $\SU(2)_1$ by $\SU(2)/\ZZ_2$. For this reason, the fusion rules of the theory are identical to those of $\SU(2)$ gauge theory. When one identifies the momentum with the conjugacy class $\mathcal{C}_\theta=\mathrm{diag}(\mathrm{e}^{2\pi i \theta},\mathrm{e}^{-2\pi i \theta})$ as in \eqref{eq:theta p relation}, the chamber function $\sigma(p_1,p_2,p_3)$ in \eqref{eq:RW sigma definition} equals $1$ precisely when the moduli space of flat $\SU(2)$ bundles
\be 
\mathcal{M}_{g,\boldsymbol{\theta}}^{\mathrm{SU}(2)}=\left\{\left.\begin{subarray}{c} A_1,\dots,A_g, \\
B_1,\dots,B_g, \\
C_1,\dots,C_n \end{subarray}\in \SU(2)\, \right|\,\prod_{i=1}^g [A_i,B_i] \prod_{j=1}^n C_j=\id\,,\ C_j \in \mathcal{C}_{\theta_j} \right\}\bigg/\SU(2) \label{eq:SU(2) moduli space}
\ee
is non-empty for $(g,n)=(0,3)$. More generally, $\mathsf{A}^\text{TQFT}_{g,n}(\boldsymbol{\theta})$ computes the volume of the moduli space $\mathcal{M}_{g,\boldsymbol{\theta}}^{\mathrm{SU}(2)}$ of flat $\SU(2)$ connections \cite{Witten:1991we,Witten:1992xu,Blau:1993hj,Cordes:1994fc}.

This realization of the Runkel-Watts theory also provides a target space interpretation as the $c=1$ string at self-dual radius with an additional $\SU(2)/\ZZ_2$ gauging, even though it is not very clear how to think about the additional gauging geometrically. We should also mention that the spectral curve that we identify below in \eqref{eq:RW spectral curve w coordinate} is tantalizingly close for $q=q'=1$ to the spectral curve of the $c=1$ string \cite{Collier:2026pxi}. The former spectral curve is up to scaling $\xx(w)=\cos(w)$ and $\yy(w)=w\sin(w)$, while the latter is $\xx(w)=\cos(w)$ and $\yy(w)=\sin(w)$.

\subsection{Examples and explicit checks} \label{subsec:examples and checks}

\paragraph{Examples.}
Let us give some examples of the amplitudes computed from the intersection number formula \eqref{eq:RW intersection number formula}. For the case of the three-point amplitude, the only contributing graph is the one with a single vertex of color $m$ and three external legs, with a trivial moduli space, and we immediately recover the first line of \eqref{eq:CLS 3pt limit 1}. 

For the torus one-point amplitude, we have two stable graphs:
\begin{equation}
\begin{gathered}
\begin{tikzpicture}[baseline={([yshift=-.5ex]current bounding box.center)},scale=.6]
    \node[shape=circle,draw=black, very thick] (A) at (0,0) {$1$};
    \draw[very thick] (A) to (0:1.3) node[right] {1};
\end{tikzpicture}
\qquad
\begin{tikzpicture}[baseline={([yshift=-.5ex]current bounding box.center)},scale=.6]
    \node[shape=circle,draw=black, very thick] (A) at (0,0) {$0$};
    \draw[very thick] (A) to (0:1.3) node[right] {1};
    \draw[very thick, out=145, in=-145, looseness=3] (A) to (A);
\end{tikzpicture}
\end{gathered}
\label{eq:RW stable graphs 11}
\end{equation}
The first graph is the smooth once-punctured torus. The only degree-one intersection numbers which enter are
\be
    \int_{\bM_{1,1}}\psi_1=\frac{1}{24}\,,\qquad
    \int_{\bM_{1,1}}\kappa_1=\frac{1}{24}\,.
\ee
It follows from \eqref{eq:RW intersection number formula} that its contribution is
\begin{align}
\sum_{m=1}^\infty
(-1)^{(q+q')m}\sin(2\pi \sqrt{qq'}\,m p_1)\,
\frac{\frac{1}{4}\left(\frac{q'}{q}+\frac{q}{q'}\right)-p_1^2}{48\pi m}\,.
\label{eq:RW 11 smooth graph}
\end{align}
The second graph is the non-separating boundary contribution. Its moduli space is $\bM_{0,3}$, so all positive-degree classes vanish. We only need the degree-zero edge factor with the same color on the two half-edges,
\be
    \frac{\Gamma(\frac{3}{2})}{\sqrt{\pi}(\pi\sqrt{qq'})^2}
    \left(-\frac{1}{(2m)^2}\right)
    =
    -\frac{1}{8\pi^2qq'm^2}\,.
\ee
Including the automorphism factor $\frac{1}{2}$ which exchanges the two half-edges of the loop, the contribution of the second graph is
\be
    -\sum_{m=1}^\infty
    (-1)^{(q+q')m}\sin(2\pi\sqrt{qq'}\,m p_1)\,
    \frac{1}{32\pi^3qq'm^3}\,.
\label{eq:RW 11 loop graph}
\ee
Adding the two graphs gives
\begin{align}
\mathsf{RW}_{1,1}(p_1)
&=
\sum_{m=1}^\infty
(-1)^{(q+q')m}\sin(2\pi\sqrt{qq'}\,m p_1)
\left[
\frac{\frac{1}{4}\big(\frac{q'}{q}+\frac{q}{q'}\big)-p_1^2}{48\pi m}
-\frac{1}{32\pi^3qq'm^3}
\right] 
\nonumber\\
&=
-\frac{1}{2}\mathsf{V}_{1,1}^{(\sqrt{q'/q})}(ip_1) B_1(\{\sqrt{qq'}\,p_1+\tfrac{q+q'}{2}\})
-\frac{1}{48qq'} B_3(\{ \sqrt{qq'}\,p_1+\tfrac{q+q'}{2}\})
\label{eq:RW 11}
\end{align}
where in the last equality we used the Bernoulli Fourier series
\begin{equation} \label{eq:Bernoulli Fourier series sine}
\sum_{m=1}^\infty\frac{\sin(2\pi m y)}{m^{2k-1}}
= (-1)^k\frac{(2\pi)^{2k-1}}{2(2k-1)!} B_{2k-1}(\{y\})\,, \quad k\in\mathbb{Z}_{\geq 1} \,,
\end{equation}
and that the VMS quantum volume $\mathsf{V}_{1,1}^{(\sqrt{q'/q})}(ip_1)
=\frac{1}{24}\big(\frac{1}{4}(\frac{q'}{q}+\frac{q}{q'})-p_1^2\big)$.

For the sphere four-point amplitude, the two topological types of stable graphs are
\begin{equation}
\begin{gathered}
\begin{tikzpicture}[baseline={([yshift=-.5ex]current bounding box.center)},scale=.6]
    \node[shape=circle,draw=black, very thick] (A) at (0,0) {$0$};
    \draw[very thick] (A) to (135:1.3) node[left] {1};
    \draw[very thick] (A) to (-135:1.3) node[left] {2};
    \draw[very thick] (A) to (-45:1.3) node[right] {3};
    \draw[very thick] (A) to (45:1.3) node[right] {4};
\end{tikzpicture}
\qquad
\begin{tikzpicture}[baseline={([yshift=-.5ex]current bounding box.center)},scale=.6]
    \node[shape=circle,draw=black, very thick] (A) at (0,0) {$0$};
    \node[shape=circle,draw=black, very thick] (B) at (-1.5,0) {$0$};
    \draw[very thick] (A) to (B);
    \draw[very thick] (B) to ++(135:1.3) node[left] {1};
    \draw[very thick] (B) to ++(-135:1.3) node[left] {2};
    \draw[very thick] (A) to (-45:1.3) node[right] {3};
    \draw[very thick] (A) to (45:1.3) node[right] {4};
    \node at (3.5,0) {$+\;2\text{ perms.}$};
\end{tikzpicture}
\end{gathered}
\label{eq:RW stable graphs 04}
\end{equation}
A similar calculation yields the following result. See appendix~\ref{app:details-of-calculations} for further details of the derivation. Define the kernel
\begin{align}
K(u,v)
&=
B_2(\{v\})
\left(
B_2(\{u+v\})+B_2(\{u-v\})
\right)
\nonumber\\
&\quad
-\frac{2}{3}B_1(\{v\})
\left(
B_3(\{u+v\})+B_3(\{v-u\})
\right)
\nonumber\\
&\quad
+\frac{1}{6}
\left(
B_4(\{u+v\})+B_4(\{u-v\})
\right)
+\frac{2}{3}B_4(\{u\})\,.
\label{eq:RW 04 K}
\end{align}
The contact graph gives
\begin{align} \label{eq:RW 04 contact result}
\mathsf{RW}^{\mathrm{con}}_{0,4}(\boldsymbol p)
&=
\frac{1}{16} \mathsf{V}_{0,4}^{(\sqrt{q'/q})}(i \boldsymbol p)
\sum_{\sigma_2,\sigma_3,\sigma_4=\pm1}
\sigma_2\sigma_3\sigma_4\,
B_2(\{x_1+\sigma_2x_2+\sigma_3x_3+\sigma_4x_4\}).
\end{align}
The exchange graph in the $12|34$ channel gives
\begin{align}
\mathsf{RW}&^{\mathrm{ex}}_{0,4;12|34}(\boldsymbol p)
\nonumber\\
&\!\!=
\frac{1}{16qq'}
\Big[
K(x_1-x_2+\tfrac{q+q'}{2},x_3-x_4+\tfrac{q+q'}{2})
-K(x_1-x_2+\tfrac{q+q'}{2},x_3+x_4+\tfrac{q+q'}{2})
\nonumber\\
&\!\!\qquad \quad
-K(x_1+x_2+\tfrac{q+q'}{2},x_3-x_4+\tfrac{q+q'}{2})
+K(x_1+x_2+\tfrac{q+q'}{2},x_3+x_4+\tfrac{q+q'}{2})
\Big] \,,
\label{eq:RW 04 exchange 1234 result}
\end{align}
where we again used the shorthand $x_j=\sqrt{qq'}\,p_j$.
The two other exchange channels are obtained by the permutations. Thus
\be
\mathsf{RW}_{0,4}(\boldsymbol p)
=
\mathsf{RW}^{\mathrm{con}}_{0,4}(\boldsymbol p)
+\mathsf{RW}^{\mathrm{ex}}_{0,4;12|34}(\boldsymbol p)
+\mathsf{RW}^{\mathrm{ex}}_{0,4;13|24}(\boldsymbol p)
+\mathsf{RW}^{\mathrm{ex}}_{0,4;14|23}(\boldsymbol p)\,.
\label{eq:RW 04 final}
\ee

\paragraph{$\mathrm{SO}(8)$ triality symmetry of $\mathsf{RW}_{0,4}(\boldsymbol p)$.}
Lastly, let us point out that the sphere four-point amplitude \eqref{eq:RW 04 final} satisfies an $\mathrm{SO}(8)$ triality symmetry, albeit with a further restriction on the momenta. 
This is a symmetry of the Liouville CFT four-point function \cite{Giribet:2009hm,Eberhardt:2023mrq} that originates from a flavor symmetry of a four-dimensional $\mathcal{N}=2$ gauge theory related to Liouville CFT via the AGT correspondence \cite{Alday:2009aq,Gaiotto:2009we}.

First, we recall that the $\mathbb{C}$LS four-point amplitude satisfies the following triality covariance property, under which the momenta transform as $p_j\to p_j-\frac{1}{2}\sum_{i=1}^4p_i$ \cite[section 3.3]{Collier:2024lys}:
\begin{equation} \label{eq:CLS triality covariance}
\mathsf{A}^{(b)}_{0,4}(\boldsymbol{p})
=
\left(
\prod_{j=1}^4
\frac{\vartheta_1(2bp_j,b^2)}{\vartheta_1(2b(p_j-\frac{1}{2} \sum_{i=1}^4 p_i),b^2)}
\right)
\mathsf{A}^{(b)}_{0,4}(\boldsymbol{p}-\tfrac{1}{2} {\textstyle\sum_{i=1}^4} p_i) \,.
\end{equation}
In the Runkel-Watts limit, $b\to \sqrt{q'/q} + i\varepsilon$ with $\varepsilon\to0$, the prefactor in \eqref{eq:CLS triality covariance} has the leading asymptotic behaviour
\begin{equation} \label{eq:prefactor triality limit}
\left|
\prod_{j=1}^4
\frac{\vartheta_1(2bp_j,b^2)}{\vartheta_1(2b(p_j-\frac{1}{2} \sum_{i=1}^4 p_i),b^2)}
\right|
\sim 
\exp \left( \frac{1}{\varepsilon}\frac{\pi}{2q\sqrt{qq'}}\Sigma(\boldsymbol{x}) \right)\,,
\end{equation}
where (using the shorthand $x_j=\sqrt{qq'}\,p_j$)
\begin{equation} \label{eq:Sigma definition}
\Sigma(\boldsymbol{x})
=\sum_{j=1}^4
\Big[
B_2\big(\{{\textstyle\sum_{i=1}^4} x_i-2x_j\}\big)-B_2(\{2x_j\})
\Big] \,.
\end{equation}
Therefore, following the same logic as in the $\mathbb{C}$LS case \cite{Collier:2024lys}, we verified that the Runkel-Watts four-point amplitude satisfies,
\begin{equation} \label{eq:RW triality invariance}
\mathsf{RW}_{0,4}(\boldsymbol{p}) = \mathsf{RW}_{0,4}(\boldsymbol{p} -\tfrac{1}{2} {\textstyle\sum_{i=1}^4} p_i) \quad \text{if }\quad \Sigma(\boldsymbol{x}) = 0 \,.
\end{equation}
Let us note that the condition $\Sigma(\boldsymbol{x}) = 0$ is a subset of the fusion rules of the RW four-point amplitude.\footnote{At the boundary of this four-point fusion rule, the RW amplitude vanishes, as can be seen for instance from the vanishing of the volume of moduli space of flat $\mathrm{SU}(2)$ connections \eqref{eq:SU(2) moduli space}.}
Furthermore, if $\Sigma(\boldsymbol{x})>0$ the triality-transformed amplitude vanishes, $\mathsf{RW}_{0,4}(\boldsymbol{p} -\tfrac{1}{2} {\textstyle\sum_{i=1}^4} p_i)=0$, whereas if $\Sigma(\boldsymbol{x})<0$ the original amplitude vanishes, $\mathsf{RW}_{0,4}(\boldsymbol{p})=0$.
It is somewhat surprising that the Runkel-Watts four-point amplitude is only invariant subject to the additional condition $\Sigma(\boldsymbol{x})=0$.\footnote{Note that the condition $\Sigma(\boldsymbol{x})=0$ includes, but is more general than, the condition that all $x_j$ together with their triality images lie in the fundamental chamber $0<x_k<\frac12,\,
0<\frac12\sum_{j=1}^4x_j-x_k<\frac12$, for $k=1,\ldots,4$.}

\paragraph{Explicit numerical integration.}
Here, we compute the Runkel-Watts amplitudes by direct numerical integration. This provides an independent check of the formulas \eqref{eq:RW 11} and \eqref{eq:RW 04 final}, which were obtained by taking the Runkel-Watts limit of the \CLS{} amplitudes. Since the calculation closely parallels that of \cite{Collier:2023cyw}, we keep the discussion brief and highlight only the main ingredients.

The torus one-point amplitude of the Runkel-Watts string can be computed directly by integrating over the moduli space of once-punctured tori:
\begin{align} \label{eq:RW 1,1 numerical_integration}
\mathsf{RW}&_{1,1}(p_1)
= \mathcal{N}_{p_1}\frac{(2\pi)^2}{2} \int_{F_0} \d^2 \tau \, |\eta(\tau)|^4 \nn
&\!\times \int_{\mathcal C} \frac{\d p}{-i} \, \rho_b(p) C_b(p_1,p,p) \mathcal F^{(b)}_{1,1}(ip_1;ip|e^{2\pi i \tau})\mathcal F^{(b)}_{1,1}(ip_1;ip|e^{-2\pi i \tau})\nn
&\!\times \int_0^\infty \!\!\!\!\d P \, \frac{P^2}{2\rho_b(P)} \big ( C_b(p_1,P,P) \big )^{-1} \sigma(p_1,P,P) \mathcal F^{(ib)}_{1,1}(p_1;P|e^{2\pi i \tau})\mathcal F^{(ib)}_{1,1}(p_1;P|e^{-2\pi i \tau}) \, .
\end{align}
Here $\mathcal F^{(b)}_{1,1}(P_1;P|e^{2\pi i \tau})$ denotes the torus one-point Virasoro conformal block at central charge $c=1+6(b+b^{-1})^2$, with external weight $h_{P_1}$ and internal weight $h_P$, where $h_P=\frac{1}{4}(b+b^{-1})^2+P^2$. The contour of the Liouville correlator is $\mathcal C= i \mathbb R_{\geq0}$. The modular integral is over the standard fundamental domain
$F_0= \{ \tau \in \mathbb C | -\frac{1}{2} \leq \text{Re}\, \tau \leq \frac{1}{2},\, |\tau|\geq 1\}$.

\begin{figure}[h]
    \centering
    \includegraphics[width=0.7\textwidth]{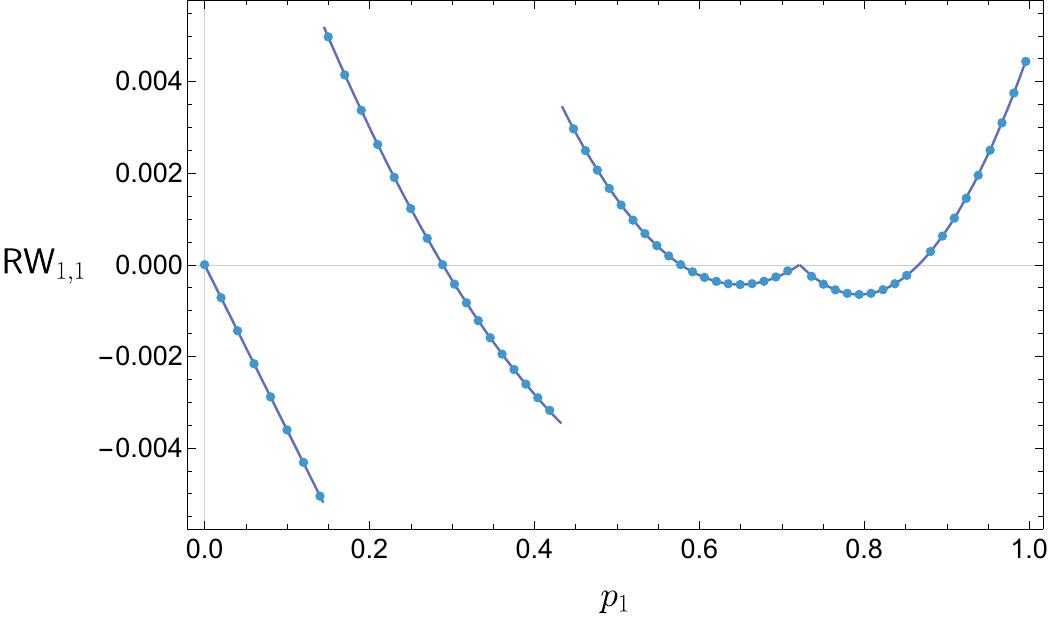}
    \includegraphics[width=0.7\textwidth]{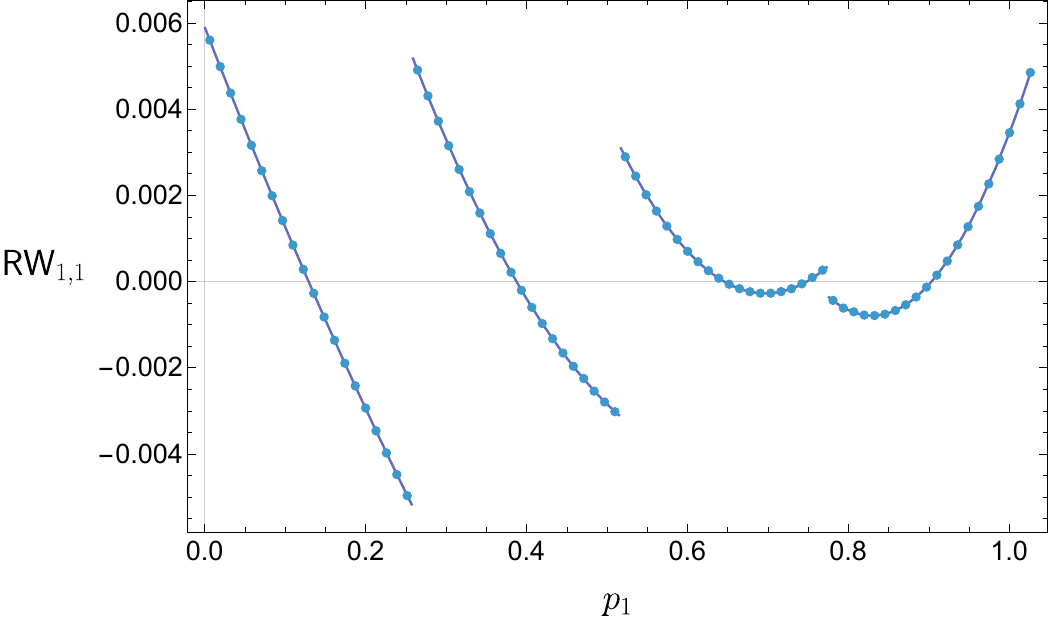}
    \caption{The torus one-point Runkel-Watts amplitude $\mathsf{RW}_{1,1}(p_1)$ as a function of $p_1$. The light blue dots are obtained from direct numerical integration, while the lavender curve is the analytic formula \eqref{eq:RW 11}. The two agree within numerical precision. \textbf{Top}: Runkel-Watts string labeled by $(4,3)$. \textbf{Bottom}: Runkel-Watts string labeled by $(5,3)$. 
    }
    \label{fig:torus1ptnumerics}
\end{figure}

For the torus one-point correlator, the relevant poles of the Liouville structure constant $C_b(p_1,p,p)$ are located at
\bal
p =\pm \frac{1}{2} p_1 \pm \frac{1}{2} \big (m+\frac{1}{2} \big ) b \pm \frac{1}{2} \big (n+\frac{1}{2} \big ) b^{-1},~~~m,n\in \mathbb Z_{\geq 0} \ .
\eal
As the external momentum is analytically continued, these poles can cross the original contour $i \mathbb R$ when their real parts change sign. Thus the condition for no pole to cross is
\bal
p_1 < \frac{b+b^{-1}}{2} \ .
\eal
In this small-momentum regime, we evaluate \eqref{eq:RW 1,1 numerical_integration} numerically. The result is shown in figure \ref{fig:torus1ptnumerics}.

\begin{figure}[h]
    \centering
    \includegraphics[width=0.7\textwidth]{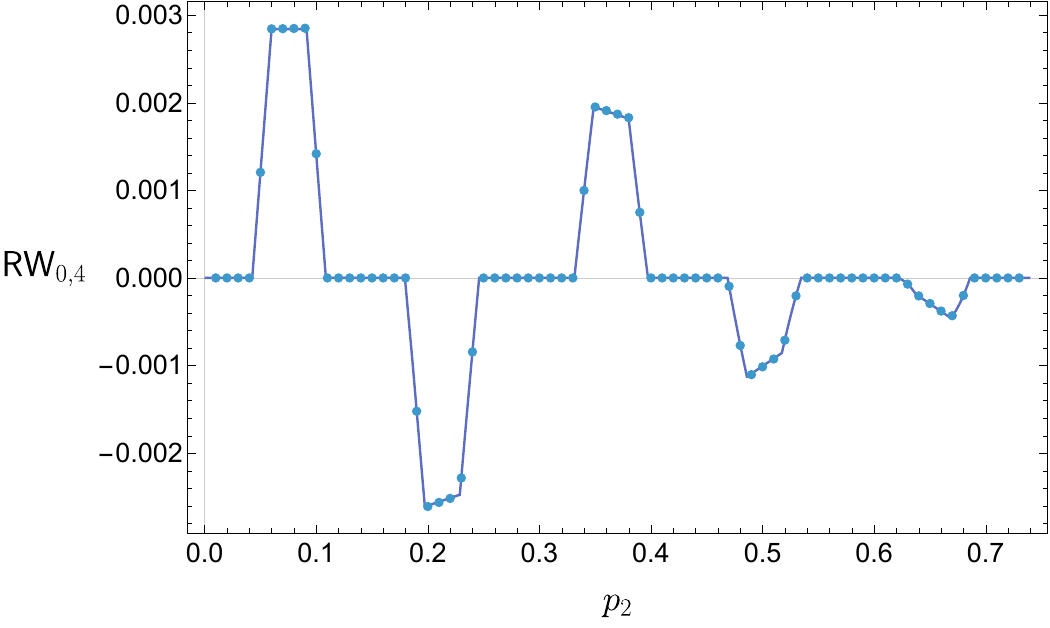}
    \includegraphics[width=0.7\textwidth]{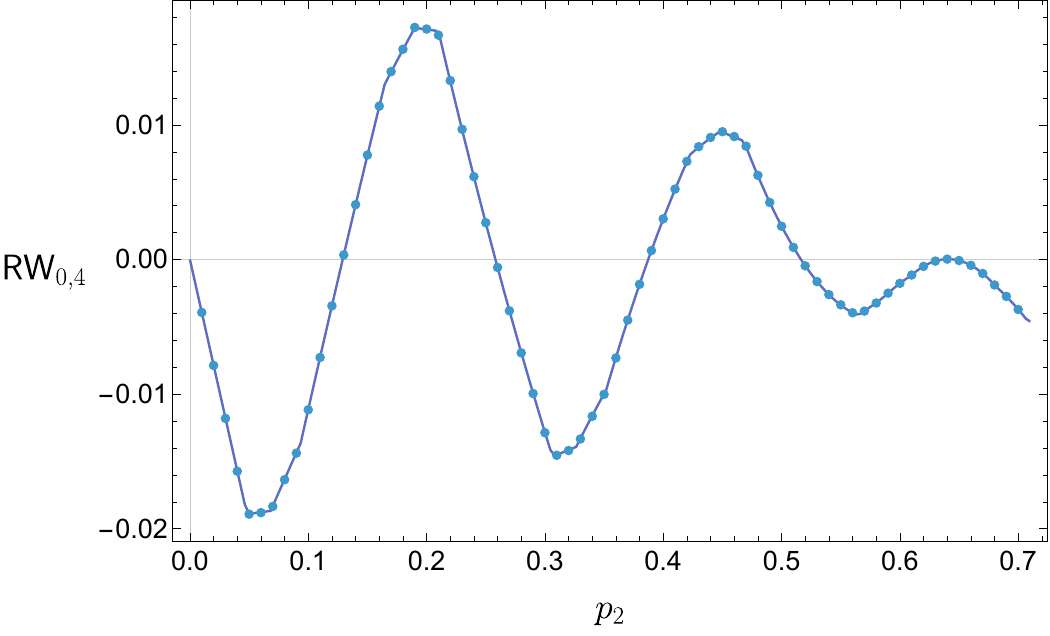}
    \caption{The sphere four-point Runkel-Watts amplitude $\mathsf{RW}_{0,4}(p_1,p_2,p_3,p_4)$ as a function of $p_2$. The light blue dots are obtained from direct numerical integration, while the lavender curve is the analytic formula \eqref{eq:RW 04 final}. The two agree within numerical precision. \textbf{Top}: Runkel-Watts string labeled by $(4,3)$, with $p_1=0.22$, $p_3=0.28$, and $p_4=0.12$. \textbf{Bottom}: Runkel-Watts string labeled by $(5,3)$, with $p_1=0.21$, $p_3=0.07$, and $p_4=0.33$.
    }
    \label{fig:sphere4ptnumerics}
\end{figure}

A direct numerical check of the sphere four-point amplitude proceeds analogously, starting from the corresponding conformal-block decomposition. Since the calculation is entirely analogous to the preceding torus one-point analysis and to that of \cite{Collier:2023cyw}, we omit the details. We again restrict to the small-momentum regime, so that no contour deformation is required. The numerical comparison is shown in figure \ref{fig:sphere4ptnumerics}.

\subsection{Properties}\label{subsec:properties}

The amplitudes of the Runkel-Watts string have a number of interesting properties. Many of them are directly inherited from the complex Liouville string and we only mention them in passing.
\paragraph{Piecewise polynomiality.} The amplitudes are all piecewise polynomial. This follows most obviously from the representation \eqref{eq:RW intersection number formula 3} where all ingredients have this property. This property is very surprising from the worldsheet point of view, where none of the ingredients are piecewise polynomial. In particular and contrary to the complex Liouville string, this also implies that the amplitudes do not have any poles. This is also surprising from the worldsheet point of view, since the integrand has various resonance poles, but the residue is a total derivative. An analogous property was demonstrated in \cite{Khromov:2025awh} from the worldsheet for the VMS volumes $\mathsf{V}_{0,4}^{(b)}$ and $\mathsf{V}_{1,1}^{(b)}$.

The amplitude \emph{does} however have many discontinuities corresponding to the piecewise polynomial behavior. In fact, the formulas that we presented correspond to a particular analytic continuation of the amplitudes outside of its initial definition range as discussed at the end of section~\ref{subsec:worldsheet-CFT}.

\paragraph{Growth conditions.} The amplitudes also only grow polynomially for large momenta and/or central charges. In fact, we have the precise statement
\be 
\mathsf{RW}_{g,n}(\boldsymbol{p}) \sim \mathsf{A}^\text{TQFT}_{g,n}(\boldsymbol{\theta}=\sqrt{qq'} \boldsymbol{p}+\tfrac{1}{2}(q+q'))\mathsf{V}_{g,n}^{(\sqrt{q'/q})}(i\boldsymbol{p})\ ,
\ee
where the asymptotics holds when we send any number of $p_i \to \infty$ and/or $c \to \infty$. $\mathsf{A}_{g,n}^\text{TQFT}$ was defined in \eqref{eq:SU(2) YM partition function}. This follows from \eqref{eq:RW intersection number formula 3}. Any stable graph with at least one edge is subleading: the edge-dependent factors are bounded, since they are Bernoulli polynomials evaluated at fractional parts, and the vertex VMS volumes have lower total polynomial degree. Thus the trivial stable graph dominates the answer, which gives the statement. Notice moreover that, for $2g-2+n\ge 2$,
\be
|\mathsf{A}^\text{TQFT}_{g,n}(\boldsymbol{\theta})| \le 2^{3-3g-n}\pi^{2-2g-n}\sum_{m=1}^\infty \frac{1}{m^{2g-2+n}}=2^{3-3g-n}\pi^{2-2g-n} \zeta(2g-2+n)
\ee
is bounded. For the base cases $2g-2+n=1$, i.e. $(g,n)=(0,3)$ and $(1,1)$, the right-hand side diverges ($\zeta(1)=\infty$), but $\mathsf{A}^\text{TQFT}_{g,n}$ is still bounded directly, being a Bernoulli polynomial evaluated at fractional parts.

\paragraph{Dilaton equation.}  The amplitudes satisfy a dilaton equation that we will discuss below in the context of topological recursion \eqref{eq:RW dilaton equation}. 
It relates $\mathsf{RW}_{g,n+1}$ for $p_{n+1}=\frac{q \pm q'}{2\sqrt{qq'}}$ to $\mathsf{RW}_{g,n}$. An analogous equation holds in the $\CC$LS. However, notice that due to the different scalings in $\varepsilon$ of the LHS and the RHS of this equation, see \eqref{eq:RW amplitude limit}, deriving such an equation as a limiting case of the $\CC$LS dilaton equation requires one to compute $\lim_{\varepsilon \to 0} \varepsilon^{2g-2+n}\mathsf{A}_{g,n+1}$ at $p_{n+1}=\frac{q \pm q'}{2\sqrt{qq'}}$. $\mathsf{RW}_{g,n+1}$ vanishes at $p_{n+1}=\frac{q \pm q'}{2\sqrt{qq'}}$, so that limit falls outside the scope of the general limit derived above. Instead, it is much more convenient to translate the general dilaton equation of topological recursion which we will do below.

\subsection{The dual spectral curve}\label{subsec:dual-spectral-curve}

As noted above, from the intersection number formula \eqref{eq:RW intersection number formula} we can extract the data of the resolvent differentials $\omega_{g,n}$, which are the main objects of interest in the topological recursion description. In particular, the seed resolvents for the Runkel-Watts string take the form
\begin{equation} \label{eq:RW seed resolvents}
\omega_{0,1}(z) = \frac{4\pi^2}{\sqrt{qq'}} 
\sin\!\left(\frac{\pi q'\sqrt{z}}{\sqrt{qq'}}\right)
\sin\!\left(\frac{\pi q\sqrt{z}}{\sqrt{qq'}}\right)\d z
\,, \quad \omega_{0,2}(z_1,z_2) = \frac{\d z_1\, \d z_2}{(z_1-z_2)^2} \,.
\end{equation}
Note that $\omega_{0,1}$ is invariant under $\sqrt{z}\to -\sqrt{z}$, and that $\omega_{0,2}$ is the standard Bergman kernel. 
The spectral curve can be parametrized by
\begin{equation} \label{eq:RW spectral curve}
\xx(z)=-2\cos\!\left(\frac{\pi \sqrt{z}}{\sqrt{qq'}}\right) \,, 
\qquad
\yy(z)=-4\pi \sqrt{z}\,
\frac{
\sin\!\left(\frac{\pi q'\sqrt{z}}{\sqrt{qq'}}\right)
\sin\!\left(\frac{\pi q\sqrt{z}}{\sqrt{qq'}}\right)
}{
\sin\!\left(\frac{\pi \sqrt{z}}{\sqrt{qq'}}\right)
} \,,
\end{equation}
upon which the seed resolvent $\omega_{0,1}(z)=-\yy(z)\d\xx(z)$.

\begin{figure}[h]
    \centering
    \includegraphics[width=0.85\textwidth]{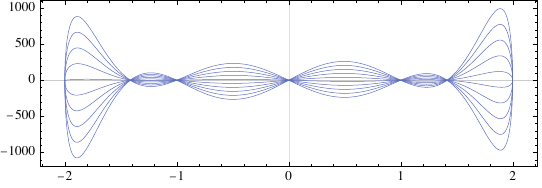}
    \caption{The spectral curve of the $(5,3)$ Runkel-Watts string, projected onto the $(\operatorname{Re} x,\operatorname{Re} y)$-plane. The portion displayed here corresponds to $z \in (0,1500)$. As one includes the remaining values of $z$, the curve continues to oscillate between the same endpoints on the $\operatorname{Re} x$-axis, with progressively larger amplitude in the $\operatorname{Re} y$ direction.}
    \label{fig:curve}
\end{figure}

\paragraph{Properties.}
The Runkel-Watts spectral curve shows an infinite number of sheets due to the invariance of $\mathsf{x}(z)$ under $\sqrt{z}\to \sqrt{z}+2\sqrt{q q'}n$ for all $n\in\mathbb{Z}$. 
Closely related are the infinitely many branch points, where two local branches meet, equivalently where $\d\mathsf{x}=0$. They are located at
\beq
z_a^\star= \left( a \sqrt{q' q}\right)^2 \quad a\in\mathbb{Z}_{>0} \,. \label{eq:branch points}
\eeq 
Near such a branch point, the curve has square-root behavior: for each nearby point $z$, there is a unique point $\sigma_a(z)$ on the other local branch with the same $\mathsf{x}$-value. This defines the local Galois involution that exchanges the two branches,
\beq
\sigma_a:\sqrt{\mathsf{x}-\mathsf{x}(z_a^\star)}\to -\sqrt{\mathsf{x}-\mathsf{x}(z_a^\star)} \,.
\eeq
The other singular points are pairs of points $z^\pm$ on the spectral curve such that
$\mathsf{x}(z^+)=\mathsf{x}(z^-)$ and $\mathsf{y}(z^+)=\mathsf{y}(z^-)$.
Their images in the $(\mathsf{x},\mathsf{y})$-plane are the nodal singularities located at
\beq
\Big\{
\Big(-2\cos\frac{\pi r}{q},0\Big)
\,\Big |\,
r=0,\ldots,q
\Big\}
\bigcup
\Big\{
\Big(-2\cos\frac{\pi s}{q'},0\Big)
\,\Big |\,
s=1,\ldots,q'-1
\Big\} \, .
\eeq
Equivalently, these are the zeroes of $\mathsf{y}(z)$. All the self-intersection points have infinite multiplicity. See figure~\ref{fig:curve} for a sample plot of (a real projection of) the spectral curve \eqref{eq:RW spectral curve}. We expect these singularities to encode instanton, or more generally non-perturbative effects. We leave their detailed study for future work.

\paragraph{Density of states.} In the context of random matrix models, we can interpret the first matrix as a Hamiltonian and define the energy $E\equiv \xx$. Since the potential term $V'_1(E)$ is analytic across the physical cut, the density of eigenvalues of the first matrix is determined by the discontinuity of $\yy(E)$. Recalling that $\omega_{0,1}(z_1)=R_{0,1}(z_1)\d \xx(z_1)$ up to analytic terms that do not contribute to the discontinuity (see \cite{Eynard:2015aea} for a review), we obtain
\beq
\rho_0(E)
=
-\frac{1}{2\pi i}\operatorname{Disc} R_{0,1}(E)
=
2\sqrt{qq'}\sqrt{E^2-4}\,
U_{q-1}\!\left(\frac{E}{2}\right)
U_{q'-1}\!\left(\frac{E}{2}\right) \ .
\eeq
The equality with the Chebyshev form holds up to orientation: the discontinuity is taken with the orientation for which $\rho_0(E)\ge 0$, which relative to the naive $E\pm\mathrm{i}0$ prescription amounts to an overall sign $(-1)^{q+q'}$. Here $U_n(z)$ denotes the $n$-th Chebyshev polynomial of the second kind, $U_n(\cos\theta)=\sin((n+1)\theta)/\sin\theta$.

\begin{figure}[t]
    \centering
    \includegraphics[width=0.7\textwidth]{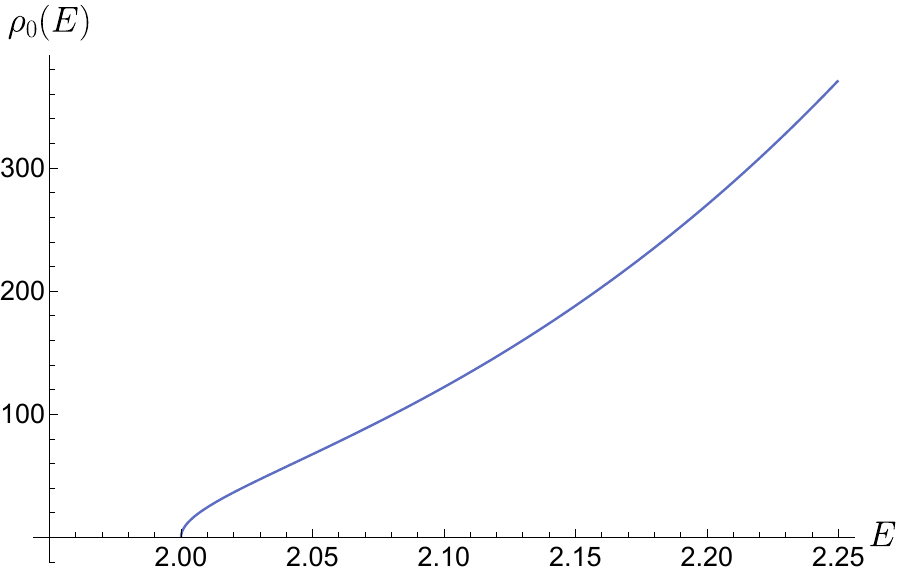}
    \caption{The density of states $\rho_0(E)$ of the $(5,3)$ Runkel-Watts string.}
    \label{fig:density of states}
\end{figure}

The density is supported on the cut starting at $E=2$. Near the edge, it behaves as $\rho_0(E)\sim 4(qq')^{3/2}\sqrt{E-2}$. At large energy, it has polynomial growth, $\rho_0(E)\sim 2(qq')^{1/2}E^{q+q'-1}$. See figure~\ref{fig:density of states} for a sample plot of the Runkel-Watts density of states.

\paragraph{Single branch point and VMS.} Let us also notice that \eqref{eq:RW seed resolvents} reduces to the VMS spectral curve if we restrict to a single branch point \eqref{eq:branch points}. This is of course not surprising in light of the appearance of the quantum volume in \eqref{eq:RW intersection number formula} that is associated to the vertices of the stable graph decomposition of the RW amplitudes. 

\paragraph{Topological recursion.} To apply topological recursion, it is convenient to work with the coordinate $w=\sqrt{z}$. In terms of $w$, the spectral curve takes the form
\begin{equation} \label{eq:RW spectral curve w coordinate}
\xx(w)=-2\cos\!\left(\frac{\pi w}{\sqrt{qq'}}\right) \,, 
\qquad
\yy(w)=-4\pi w\,
\frac{
\sin\!\left(\frac{\pi q'w}{\sqrt{qq'}}\right)
\sin\!\left(\frac{\pi qw}{\sqrt{qq'}}\right)
}{
\sin\!\left(\frac{\pi w}{\sqrt{qq'}}\right)
} \,,
\end{equation}
together with 
\begin{equation}
\omega_{0,1}(w)=8\pi^2 \,\frac{w}{\sqrt{qq'}}\,
\sin\!\left(\frac{\pi q' w}{\sqrt{qq'}}\right)
\sin\!\left(\frac{\pi q w}{\sqrt{qq'}}\right)\,\d w \,. \label{eq:omega01 w coordinates}
\end{equation}
The Bergman kernel takes the form
\begin{equation}
\omega_{0,2}(w_1,w_2)=\left[\frac{1}{(w_1-w_2)^2}-\frac{1}{(w_1+w_2)^2}\right]\,\d w_1\d w_2 \,.
\end{equation}
The branch points are located at $w_m^*=\sqrt{qq'}\,m$, with $m\in\mathbb{Z}_{\geq 1}$, and the local Galois involution at the $m^{\mathrm{th}}$ branch point can be taken to be $\sigma_m(w)=2\sqrt{qq'}\,m-w$. 
In these coordinates, the recursion kernel is
\begin{align}
K_m(w_1,w)=
\frac{
w_1\left(\frac{1}{w_1^2-w^2}-\frac{1}{w_1^2-\sigma_m(w)^2}\right)
}{
16\pi^2 m\,
\sin\!\left(\frac{\pi q'}{\sqrt{qq'}}w\right)\sin\!\left(\frac{\pi q}{\sqrt{qq'}}w\right)
}
\frac{\d w_1}{\d w}\,,
\end{align}
and the general topological recursion formula reads
\begin{multline}
    \omega_{g,n}(w_1,\boldsymbol{w})=\sum_{m\text{ branch pts}}\Res_{w=w^*_m} K_m(w_1,w) \bigg(\omega_{g-1,n+1}(w,\sigma_m(w),\boldsymbol{w})\\
    +\sum_{h=0}^g \sum_{{\substack{I\cup I^c=\{2,\ldots,n\}\\ \{h,I\}\neq \{0,\emptyset\}\\ \{h,I^c\}\neq \{g,\emptyset\}}}} \omega_{h,1+|I|}(w,\boldsymbol{w}_{I}) \, \omega_{g-h, 1+|I^c|}(\sigma_m(w),\boldsymbol{w}_{I^c})\bigg)\,.\label{eq:TR}
\end{multline}

\paragraph{Examples.}
As simple examples, one finds the following resolvent differentials. For the tree-level three-point resolvent,
\begin{align}\label{eq:TR omega03}
   \omega_{0,3}(w_1,w_2,w_3)
   &=
   \sum_{m=1}^\infty
   \Res_{w=w_m^*}
   K_m(w_1,w)
   \Big(
      \omega_{0,2}(w,w_2)\omega_{0,2}(\sigma_m(w),w_3)
      +(w_2\leftrightarrow w_3)
   \Big)\nonumber\\
   &=
   -\sum_{m=1}^\infty
   \frac{8\,(qq')^{\frac{3}{2}} m^2\,(-1)^{(q+q')m}\,w_1w_2w_3\,\d w_1 \d w_2 \d w_3}
   {\pi^4\,(w_1^2-(w_m^*)^2)^2(w_2^2-(w_m^*)^2)^2(w_3^2-(w_m^*)^2)^2},
\end{align}
while for the torus one-point resolvent,
\begin{align}\label{eq:TR omega11}
    \omega_{1,1}(w_1) &=
    -\sum_{m=1}^\infty
    \frac{\sqrt{qq'}\,(-1)^{(q+q')m}\,w_1\,\d w_1}
    {96\pi^4\,(w_m^*)^2(w_1^2-(w_m^*)^2)^4}
    \Bigg(
      24(w_m^*)^4 + 12(w_m^*)^2\big(w_1^2-(w_m^*)^2\big)\nonumber\\
    &\qquad\qquad\qquad\qquad
    +\left(-6+\pi^2\big(\tfrac{q'}{q}+\tfrac{q}{q'}\big)(w_m^*)^2\right)\big(w_1^2-(w_m^*)^2\big)^2
    \Bigg)\, .
\end{align}
The overall minus signs again arise from the fact that $\d \sigma_m(w)=-\d w$.
Similarly, for the tree-level four-point differential, we have 
\begin{align}
\label{eq:TR omega04}
\omega_{0,4}(w_1 &,w_2,w_3,w_4)=\sum_{\substack{m_1,m_2=1\\m_1\neq m_2}}^\infty \frac{ (q q') m_1^2m_2^2 (-1)^{(q+q')(m_1+m_2)}}{\pi^8(m_1^2-m_2^2)^2}
\nonumber\\
&\quad \times \frac{16\, w_1 w_2 w_3 w_4\, \d w_1  \d w_2 \d w_3 \d w_4}{(w_1^2-(w_{m_1}^*)^2)^2(w_2^2-(w_{m_1}^*)^2)^2(w_3^2-(w_{m_2}^*)^2)^2(w_4^2-(w_{m_2}^*)^2)^2}
\nonumber\\
&+\sum_{m=1}^\infty  \left(\frac{m^2(q q')^2\left(\frac{q}{q'}+\frac{q'}{q}\right)}{24\pi^6} -\frac{q q'}{16 \pi^8}+\frac{q q'\mathsf{P}(w_1,w_2,w_m^*)}{16\pi^8(w_1^2-(w_{m}^*)^2)^2(w_2^2-(w_{m}^*)^2)^2}\right)
\nonumber\\
&\quad \times \frac{16\, w_1 w_2 w_3 w_4\, \d w_1  \d w_2 \d w_3 \d w_4}{(w_1^2-(w_{m}^*)^2)^2(w_2^2-(w_{m}^*)^2)^2(w_3^2-(w_{m}^*)^2)^2(w_4^2-(w_{m}^*)^2)^2}
\nonumber\\
&+\text{ 2 other cyclic perms. of \{2,3,4\}}\,, \vphantom{\int}
\end{align}
where 
\begin{align}
\mathsf{P}(w_1,w_2,w_m^*)&=24 (w_m^*)^2(w_1^2-(w_m^*)^2)^2\left((w_2^2-(w_m^*)^2)+2(w_m^*)^2\right) 
\nonumber\\
&\quad +8(w_m^*)^2\left(w_2^2-(w_m^*)^2\right)^2\left(w_1^2+(w_m^*)^2\right)\,.
\end{align}

\paragraph{Dictionary.} To recover the Runkel-Watts string amplitudes from the resolvent differentials, we use the same transform as in the $\mathbb{C}$LS or $c=1$ cases:
\begin{align} \label{eq:dictionary 1}
\mathsf{RW}_{g,n}(p_1,\dots,p_n)
&=\sum_{m_1,\dots,m_n=1}^\infty  \Res_{z_1=z_{m_1}^*} \cdots \Res_{z_n=z_{m_n}^*} \prod_{j=1}^n \frac{\cos(2 \pi p_j \sqrt{z_j})}{p_j}\,  \omega_{g,n}(z_1,\dots,z_n) 
\nonumber\\
&=\sum_{m_1,\dots,m_n=1}^\infty  \Res_{w_1=w_{m_1}^*} \cdots \Res_{w_n=w_{m_n}^*} \prod_{j=1}^n \frac{\cos(2 \pi p_j w_j)}{p_j}\,  \omega_{g,n}(w_1,\dots,w_n) 
\nonumber\\
&=\left( \int_{\mathbb{R}-i\epsilon} - \int_{\mathbb{R}+i\epsilon} \right) \prod_{j=1}^n \frac{\e^{2\pi i p_jw_j}}{4\pi i p_j} \omega_{g,n}(w_1,\dots,w_n) \,.
\end{align}
In the last line we have opened up the residue contours into the difference of horizontal contours just above and below the real axis; convergence of these contours requires the momenta $p_j$ to be real-valued. 
In practice, it is most convenient to use the RHS of the second equality to recover the RW amplitude. 
As in other minimal strings, we may interpret this dictionary as a discrete Fourier transform from position to momentum space.

The inverse dictionary is given by 
\begin{align} \label{eq:inverse dictionary}
\omega_{g,n}(w_1,\dots,w_n) = \int_0^\infty \prod_{j=1}^n \big(-4\pi i p_j \d p_j {\e}^{\pm 2\pi i p_j w_j} \big) \mathsf{RW}_{g,n}(\pm p_1,\dots,\pm p_n) \,,
\end{align}
where the $\pm$ signs in the exponential are correlated with the corresponding signs in the argument of the amplitude.

\paragraph{Recovering Runkel-Watts amplitudes.}
Applying \eqref{eq:dictionary 1} to the tree-level three-point resolvent \eqref{eq:TR omega03} gives
\begin{align}
\mathsf{RW}_{0,3}(p_1,p_2,p_3)
= \sum_{m=1}^\infty \frac{(-1)^{m(q'+q)} \prod_{j=1}^3\sin(2\pi m \sqrt{qq'}\, p_j)}{m \pi} \,,
\end{align}
which reproduces the three-point amplitude in the first equality of \eqref{eq:CLS 3pt limit 1}.
Similarly, applying \eqref{eq:dictionary 1} to the genus-one one-point resolvent \eqref{eq:TR omega11} gives
\begin{align}
\mathsf{RW}_{1,1}(p_1)    
= \sum_{m=1}^\infty
\frac{(-1)^{m(q'+q)}\sin(2\pi m \sqrt{qq'}\, p_1)}{2\pi m}
\left(
\mathsf{V}_{1,1}^{(\sqrt{q'/q})}(ip_1)-\frac{1}{16\pi^2 qq' m^2}
\right) \,,
\end{align}
thereby reproducing the amplitude in the first equality of \eqref{eq:RW 11}. We can apply the dictionary to \eqref{eq:TR omega04} and recover the Runkel-Watts string amplitude at four-point 
\begin{align}
 \mathsf{RW}_{0,4}(p_1 &,p_2,p_3,p_4)
 \nonumber\\
&=\sum_{m=1}^{\infty}\left(\frac{1}{4}\frac{\left(\frac{q}{q'}+\frac{q'}{q}\right)}{6\pi^2 m^2}-\frac{1}{16\pi^4 q q' m^4 }-\frac{\frac{p_1^2}{3}+p_2^2}{2\pi^2 m^2 }\right)\prod_{j=1}^{4}\sin(2\pi m\sqrt{q q'}p_j)
\nonumber\\
&\quad +\sum_{\begin{subarray}{c} m_1,m_2=1 \\ m_1 \ne m_2 \end{subarray}}^\infty \frac{(-1)^{(q+q')(m_1+m_2)}}{q q'\pi^4(m_1^2-m_2^2)^2}\prod_{i=1}^2\sin(2\pi m_1\sqrt{q q'}p_i)\prod_{j=3}^4\sin(2\pi m_2\sqrt{q q'}p_j)
\nonumber\\
&\quad +\text{ 2 other cyclic perms. of \{2,3,4\}}\,, \vphantom{\int}
\end{align}
which reproduces the amplitude in \eqref{eq:RW 04 final}; for instance, the second term in the first line is the diagonal component of the color sum in the exchange graph $\mathsf{RW}^{\mathrm{ex}}_{0,4;12|34}(\boldsymbol p)$ (c.f. \eqref{eq:RW 04 exchange color appendix}).
\paragraph{Mirzakhani recursion for $\mathsf{RW}_{g,n}(\boldsymbol{p})$.}
The topological recursion formula \eqref{eq:TR} can be translated directly into a recursion relation for the Runkel-Watts amplitudes $\mathsf{RW}_{g,n}$ via the inverse dictionary \eqref{eq:inverse dictionary}.
This recursion is a direct analogue of Mirzakhani's recursion for the Weil-Petersson volumes, now applied to the Runkel-Watts string amplitudes.
The derivation is completely analogous to the corresponding manipulations in the $\mathbb{C}$LS and $c=1$ cases.
We will therefore not write the explicit formula here, but simply note that a three-term recursion exists and follows directly from \eqref{eq:TR}.

\paragraph{Dilaton equation.}
In topological recursion, the differentials $\omega_{g,n}$ satisfy the so-called dilaton equation \cite{Eynard:2007kz}:
\begin{equation} \label{eq:topological recursion dilaton equation}
    \sum_{m \text{ branch pts}} \Res_{z_{n+1}=z_m^*} F_{0,1}(z_{n+1}) \, \omega_{g,n+1}(\boldsymbol{z},z_{n+1}) = (2-2g-n) \, \omega_{g,n}(\boldsymbol{z}) \,,
\end{equation}
where $\d F_{0,1}=\omega_{0,1}$. 
We can use the dictionary \eqref{eq:dictionary 1} to translate this into a dilaton equation for the Runkel-Watts amplitudes $\mathsf{RW}_{g,n}$.
In this case, using that seed resolvent $\omega_{0,1}$ is given by \eqref{eq:omega01 w coordinates}, we have
\begin{equation} \label{eq:RW F01}
    F_{0,1}(w) = \frac{1}{\sqrt{qq'}}\left[ \partial_p \frac{\cos(2\pi p w)}{p}\biggr|_{p=\frac{q+q'}{2\sqrt{qq'}}} - \partial_p \frac{\cos(2\pi p w)}{p}\biggr|_{p=\frac{q-q'}{2\sqrt{qq'}}} \right] \,.
\end{equation}
Apart from the derivative, and together with the sum over residues at all branch points in \eqref{eq:topological recursion dilaton equation}, this is exactly the same as the dictionary \eqref{eq:dictionary 1} for the Runkel-Watts amplitudes. 
Therefore, this leads to the following dilaton equation for the Runkel-Watts amplitudes:
\begin{align} \label{eq:RW dilaton equation}
    \partial_p \mathsf{RW}_{g,n+1}(\boldsymbol{p},p) \Big|_{p=\frac{q+q'}{2\sqrt{qq'}}} - \partial_p \mathsf{RW}_{g,n+1}(\boldsymbol{p},p) \Big|_{p=\frac{q-q'}{2\sqrt{qq'}}}
    =-\sqrt{qq'}(2g-2+n) \mathsf{RW}_{g,n}(\boldsymbol{p}) \,.
\end{align}
It is interesting to note that, in contrast to the analogous dilaton equations for the $\mathbb{C}$LS, VMS, and $c=1$ strings, \eqref{eq:RW dilaton equation} requires taking a derivative before setting $p=\tfrac{1}{2}(b^{-1}\pm b)$.
The derivative is essential: in the representation \eqref{eq:RW intersection number formula}, directly evaluating the $(n+1)^{\mathrm{th}}$ external leg factor at these special momenta vanishes, since $\sin(2\pi \sqrt{qq'}\,m_{n+1} p_{n+1})=\sin(\pi m_{n+1} (q\pm q'))=0$. 
In appendix~\ref{app:details-of-calculations}, we explicitly check the dilaton equation \eqref{eq:RW dilaton equation} in the simplest nontrivial case of $\mathsf{RW}_{0,4}\to\mathsf{RW}_{0,3}$.

%%%%%%%%%%%%%%%%%%%%%%%%%%%%%%%%%%%%%%%%%%%%%%%%%%%%%%%%%%%%%%%%
\section{Relation to other minimal strings}\label{sec:relation-to-other-minimal-strings}

\subsection{A factorizing VMS limit}\label{subsec:factorized-VMS}

\paragraph{Factorized YM$\otimes$VMS limit.} An interesting simplifying limit of the Runkel-Watts string is obtained by taking $q,q'\to\infty$ while keeping their ratio fixed, so that $\sqrt{q'/q}$ approaches an arbitrary real value $b_0\in\mathbb{R}$.  
In this limit, the intersection theory formula reduces to a single stable graph with a single vertex and $n$ external legs, and the edge factors become trivial. Indeed, each edge is suppressed as $\frac{1}{qq'}$ and will thus disappear from the equation in \eqref{eq:RW intersection number formula 3}.
Thus we find in the limit $q,q' \to \infty$ with $\sqrt{q'/q} \to b_0$,
\begin{align}\label{eq:coarse-grained VMS limit}
\mathsf{RW}_{g,n}(\boldsymbol{p}) \sim \mathsf{A}_{g,n}^\text{TQFT}(\sqrt{qq'} \boldsymbol{p}+\tfrac{1}{2}(q+q')) \mathsf{V}_{g,n}^{(b_0)} (i\boldsymbol{p})~ .
\end{align}
Thus, in this limit, the amplitudes are proportional to the VMS quantum volumes. 
However, the $\SU(2)$ Yang-Mills partition function $\mathsf{A}_{g,n}^\text{TQFT}$ is a highly oscillatory function of the momenta, and does not have a well-defined limit. 

\begin{figure}[t]
    \centering
    \includegraphics[width=0.7\textwidth]{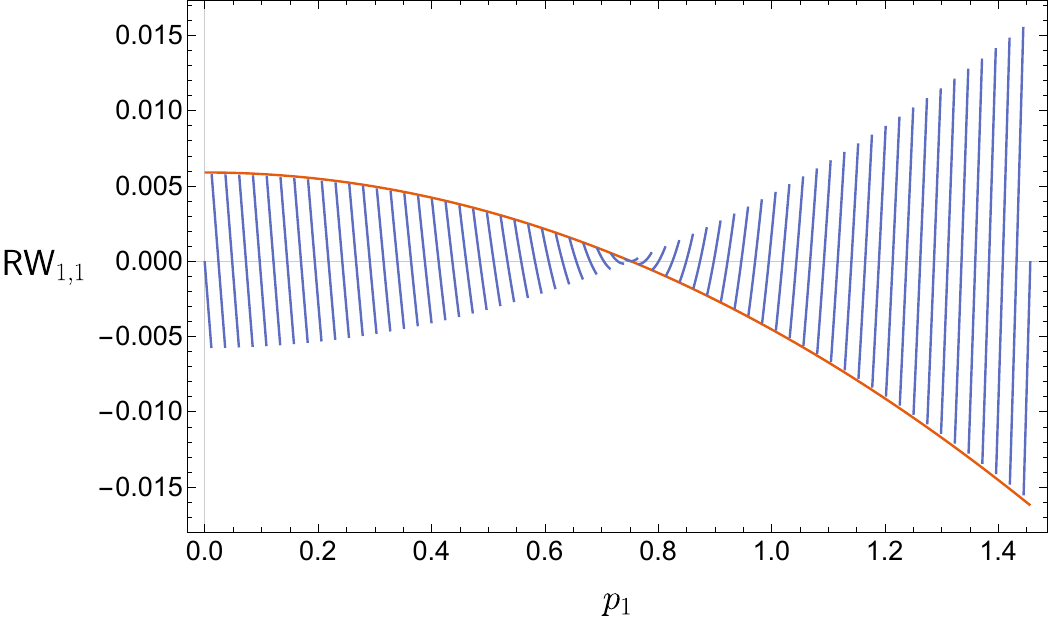}
    \caption{The Runkel-Watts amplitudes $\mathsf{RW}_{g,n}(\boldsymbol{p})$ in the limit $q,q'\to\infty$ with $\sqrt{q'/q}\to b_0\in\mathbb{R}$ held fixed are rapidly oscillating functions of the momenta whose envelope is given by the VMS quantum volume $\pm\mathsf{V}^{(b_0)}_{g,n}(i\boldsymbol{p})$. The blue curve shows the torus one-point RW amplitude for the model labeled by $(q,q')=(53,32)$, while the red curves show the corresponding VMS volume envelope.
    }
    \label{fig:coarse-grained VMS torus 1pt}
\end{figure}

In a distributional sense, when averaging over the momenta $p_j$ over a unit cell of size $1/\sqrt{qq'}$, this factor averages to zero.
Therefore, the irrational central charge limit of the Runkel-Watts string is a \emph{coarse-grained} version of the VMS quantum volumes: a highly oscillating function of the momenta whose envelope is given by the VMS quantum volumes, and which averages to zero. This is depicted in figure \ref{fig:coarse-grained VMS torus 1pt} for the torus one-point amplitude. This limit roughly realizes the expectation on the worldsheet that timelike Liouville theory can be obtained as a similar limit of the Runkel-Watts theory \cite[eq. (3.2.39)]{Ribault:2014hia}.

\paragraph{Coarse graining.} We can also make the distributional average nonzero while preserving proportionality to the VMS quantum volumes by dividing by the normalization factor $\sin(2\pi \theta)$ that would naturally lead to the standard $\SU(2)$ Yang-Mills partition function as in \eqref{eq:SU(2) YM partition function}. Noting that
\be 
\overline{\chi_m^{\mathrm{SU}(2)}(\theta)}=\int_0^1 \d \theta \chi_m^{\mathrm{SU}(2)}(\theta)=\delta_{m,2\ZZ+1}\ ,
\ee
since only odd-dimensional representations possess a charge-neutral state, we have the average
\be 
\overline{\sum_{m=1}^\infty \frac{\prod_{i=1}^n \chi_m^{\SU(2)}(\theta_i)}{m^{2g-2+n}}}=\sum_{m=1\, \text{$m$ odd}}^\infty \frac{1}{m^{2g-2+n}}=(1-2^{2-2g-n}) \zeta(2g-2+n)\ .
\ee
Thus the coarse-grained amplitudes obtained after removing the RW normalization factors are again proportional to the VMS quantum volumes.\footnote{Note that this is distinct from the simpler limit obtained by zooming in on a single branch point of the spectral curve, which again reduces to the VMS curve.}

\paragraph{Consequences for VMS.} The fact that we can extract the bare VMS quantum volume from the RW string means that the VMS quantum volumes also inherit some of the properties of the RW amplitudes discussed in section~\ref{subsec:properties}. In particular, it inherits the dilaton equation \eqref{eq:RW dilaton equation}. However, the corresponding equation does not have derivatives because of the different normalization of vertex operators that does not vanish for $p=\frac{q\pm q'}{2\sqrt{qq'}}$. 

This observation also opens up the possibility of deriving various other unknown CFT data of timelike Liouville from ordinary Liouville theory. We find it intriguing that e.g.\ the Virasoro fusion kernel is only satisfactorily known for rational $b^2$ in the $c\le 1$ regime \cite{Roussillon:2025tmv}, which corresponds exactly to the values appearing in Runkel-Watts CFT.

\subsection{A-series minimal model to Runkel-Watts-type CFT}\label{subsec:A-series-to-RW-CFT}

The original example of Runkel-Watts-type CFT, the $c=1$ case, was first discovered as the $c=1$ limit of A-series minimal model \cite{Runkel:2001ng}. We will follow the convention in \cite{Ribault:2024rvk}.

\paragraph{A-series minimal model.}

We briefly recall the A-series minimal model (AMM) conventions that will be used below. 
The A-series minimal model is labeled by two coprime positive integers $(\mathfrak q,\mathfrak q')$, with
\bal
c=1-6\frac{(\mathfrak q-\mathfrak q')^2}{\mathfrak q \mathfrak q'}\ .
\eal
Its spectrum is diagonal,
\bal
\mathcal S^{A}_{\mathfrak q,\mathfrak q'}
=
\frac{1}{2}
\bigoplus_{r=1}^{\mathfrak q-1}
\bigoplus_{s=1}^{\mathfrak q'-1}
\mathcal R_{r,s}\otimes \overline{\mathcal R}_{r,s},
\qquad 
(r,s)\sim (\mathfrak q-r,\mathfrak q'-s)\ ,
\eal
and we denote the corresponding primary by $V_{r,s}$. Its conformal weights are
\bal
h_{r,s}=\bar h_{r,s}
=\frac{c-1}{24}+P_{r,s}^2,
\qquad 
P_{r,s}
=\frac{r \mathfrak q'-s \mathfrak q}{2\sqrt{\mathfrak q \mathfrak q'}} \ .
\eal
The fusion rules are
\beq\label{eq:MMfusionrule}
V_{r_1,s_1} \times V_{r_2,s_2} = \sum_{r_3\overset{2}{=}|r_1-r_2|+1}^{\text{min}(r_1+r_2,2\mathfrak q-r_1-r_2)-1} \sum_{s_3\overset{2}{=}|s_1-s_2|+1}^{\text{min}(s_1+s_2,2\mathfrak q'-s_1-s_2)-1} V_{r_3,s_3} \ .
\eeq
where the notation $\overset{2}{=}$ means that the summation variable is increased in steps of two. Equivalently, we write the fusion rule as an indicator $f_{r_1,s_1;r_2,s_2;r_3,s_3}\in \{0,1\}$.

In the normalization used in this paper, the sphere three-point function is
\beq\label{eq:MMstructureconstant}
C_{r_1,s_1;r_2,s_2;r_3,s_3} = f_{r_1,s_1;r_2,s_2;r_3,s_3}\times  C_{\sqrt{\mathfrak q/\mathfrak q'}}(P_{r_1,s_1},P_{r_2,s_2},P_{r_3,s_3})^{-1} \ .
\eeq

\paragraph{Runkel-Watts CFT from minimal models.} To obtain the Runkel-Watts CFT labeled by $(q,q')$, we choose a sequence of A-series minimal models labeled by $(\mathfrak q,\mathfrak q')$ such that
\beq
\mathfrak q' q-\mathfrak q q' = 1,\qquad \mathfrak q,\mathfrak q' \in \mathbb Z_{>0} \ .
\eeq
This condition implies that $\mathfrak q$ and $\mathfrak q'$ are coprime. Moreover, as $\mathfrak q,\mathfrak q'\to\infty$, their ratio approaches a fixed value,
\beq
\lim_{\mathfrak q\rightarrow \infty} \frac{\mathfrak q'}{\mathfrak q} = \frac{q'}{q} \ .
\eeq
Thus, the central charge of the A-series minimal model approaches that of the desired RW-CFT. At the same time, the discrete AMM spectrum becomes dense and approaches a continuous spectrum on $\mathbb R$. For a given primary in the AMM spectrum labeled by $(r,s)$, we have
\beq
P_{r,s} = \frac{r\mathfrak q'-s\mathfrak q}{2\sqrt{\mathfrak q\mathfrak q'}}=\frac{1}{2\sqrt{qq'}} \Big ( rq'-sq + \frac{r +s\frac{q}{q'}}{2\mathfrak q} + \mathcal O(\mathfrak q^{-2}) \Big) \ .
\eeq
Since $\mathfrak q\rightarrow \infty$, the finite nontrivial part of the AMM spectrum comes from labels $r,s$ that both scale linearly with $\mathfrak q$, while $rq'-sq$ remains finite. In this regime,
\bal
\lim_{\mathfrak q\rightarrow \infty} P_{r,s} = \frac{n+x}{2\sqrt{qq'}} , \qquad n= rq'-sq  \in \mathbb Z,~~~x = \lim_{\mathfrak q\rightarrow \infty} \frac{r}{\mathfrak q} = \lim_{\mathfrak q\rightarrow \infty} \frac{sq}{q'\mathfrak q} \ .
\eal
We therefore see that, in the limit, the AMM spectrum fills the whole real line $P\in\mathbb R$, giving precisely the spectrum of the RW-CFT.

The next check concerns the fusion rules of the A-series minimal model, which in the RW limit naturally reproduce the $\sigma$ function. We briefly sketch the argument. The fusion of A-series minimal model is given by \eqref{eq:MMfusionrule}. We now take the RW limit described above. Choose sequences $(r_1,s_1)$ and $(r_2,s_2)$ that approach the momenta labeled by $(n_1,x_1)$ and $(n_2,x_2)$, respectively. First, the resulting integer part $n_3=r_3 q' - s_3 q$ becomes unbounded, ranging over either $2\mathbb Z$ or $2\mathbb Z+1$, depending on $n_1$, $n_2$, $q$, and $q'$. Second, the fractional parts become dense. Since $r_3 \rightarrow \mathfrak q x_3$, and both $r_1$ and $r_2$ scale linearly with $\mathfrak q$, we find $x_3 \in (|x_1-x_2|,\text{min}(x_1+x_2,2-x_1-x_2))$.
A nontrivial point is that the resulting values of $P_3$ are not merely dense, but are equidistributed in these intervals. Together, these facts give precisely the $\sigma(P_1,P_2,P_3)$ function of RW-CFT.

As a consequence, the RW limit of the A-series minimal-model structure constant \eqref{eq:MMstructureconstant} reproduces the RW-CFT structure constant. No additional subtlety arises from the inverse DOZZ factor: it is analytic in this limit, and is insensitive to the details of how the RW limit is taken.

The checks of the spectrum and the structure constants then establish the relation between AMM and RW-CFT correlators, for arbitrary genus and arbitrary number of punctures, via conformal block expansions.

\paragraph{Implication for the minimal string.} This discussion implies that we can obtain the RW string also as a limit of the A-series $(\mathfrak q,\mathfrak q')$-minimal string (AMS) by taking these limits directly on the level of the string amplitudes. We thus obtain the network of minimal string theories and their relation in figure~\ref{fig:minimal string theories}.

The torus one-point amplitude of $\mathrm{AMS}_{\mathfrak q,\mathfrak q'}$ was proposed in e.g. \cite{Collier:2024kwt,Rodriguez:2025rte}. Choosing the odd/odd representative $r,s\in 2\ZZ+1$, it is given by
\beq
T_{r,s}^{(\mathfrak q,\mathfrak q')} = (\mathfrak q-r)(\mathfrak q'-s) (2\mathfrak q\mathfrak q'-r\mathfrak q'-s\mathfrak q) -3 \sum_{m\overset{2}{=}1-\mathfrak q+r}^{\mathfrak q-r-1} \sum_{n\overset{2}{=}1-\mathfrak q'+s}^{\mathfrak q'-s-1} \left|m\mathfrak q'+n\mathfrak q\right| \, , \label{eq:AMS-torus-one-point}
\eeq
Taking the RW limit described above, and fixing the overall normalization appropriately, we find
\begin{align} \lim_{\mathfrak q,\mathfrak q'\to\infty} \frac{T_{r,s}^{(\mathfrak q,\mathfrak q')}}{192\,\mathfrak q\mathfrak q'} &= \mathsf{RW}_{1,1}(p_1) \, ,
\label{eq:AMS-torus-one-point-RW-limit} \end{align}
which is precisely the Runkel-Watts torus one-point amplitude \eqref{eq:RW 11}\footnote{For the odd/odd representative, $n=rq'-sq\equiv q'-q\pmod 2$, so
$\sqrt{qq'}\,p_1+\frac{q+q'}{2}\in\mathbb Z+\frac{x}{2}$. Since $x\in(0,1)$, the momentum-dependent sign factor in $\mathcal N_{p_1}$ is therefore $\operatorname{sgn}\sin(\pi x)=+1$.}. We sketch the proof in appendix~\ref{app:details-of-calculations}. 

\begin{figure}[t]
    \centering
    \includegraphics[width=0.7\textwidth]{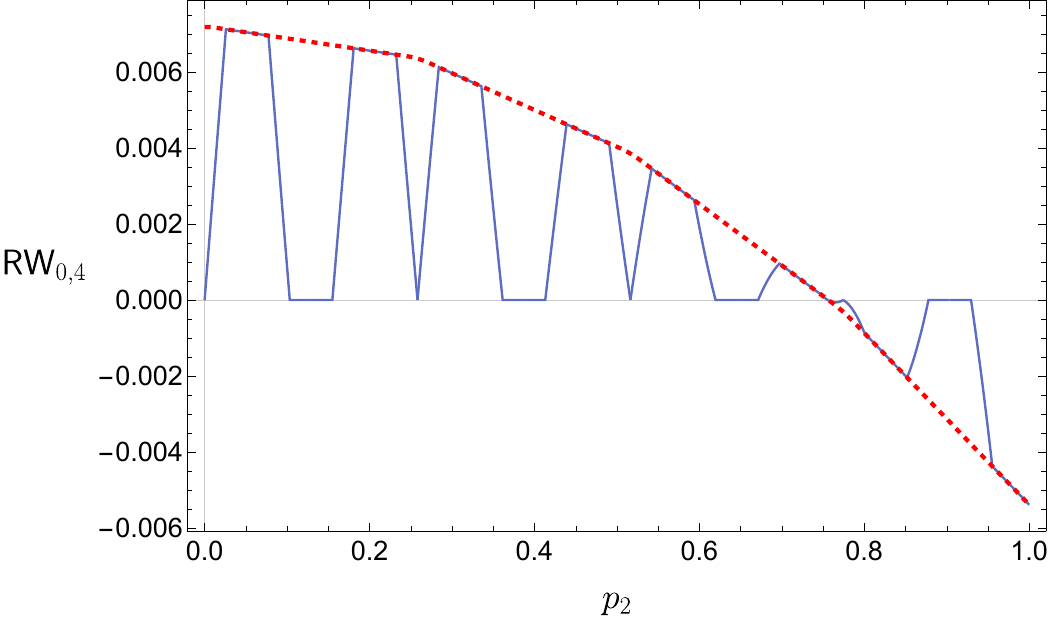}
    \caption{The lavender curve shows the RWS sphere four-point amplitude $\mathsf{RW}_{0,4}(p_1,p_2,p_3,p_4)/(\prod_{j=1}^4\mathcal{N}_{p_j})$ as a function of the external momentum $p_2$, while the red dashed curve shows the appropriately rescaled Runkel-Watts limit of the Belavin-Zamolodchikov formula. Here the RWS is labeled by $(q,q')=(5,3)$, with $p_1=0.0129$, $p_3=0.0387$, and $p_4=0.0516$, while the AMS is labeled by $(\mathfrak q,\mathfrak q')=(50003,30002)$.}
    \label{fig:RWlimit_AMS_sphere_4pt}
\end{figure}

Finally, we comment on the Belavin-Zamolodchikov formula \cite{Belavin:2006ex} for the AMS sphere four-point amplitude,
\beq
\mathrm N_{0,4}^{(b),\mathrm{BZ}}
=
r_1s_1(r_1\mathfrak q'+s_1\mathfrak q)
-
\sum_{i=2}^4
\sum_{r\overset{2}{=}1-r_1}^{r_1-1}
\sum_{s\overset{2}{=}1-s_1}^{s_1-1}
\big|
(r_i-r)\mathfrak q'-(s_i-s)\mathfrak q
\big|
\, .
\eeq
The Runkel-Watts limit of this expression does not reproduce the full RWS sphere four-point amplitude for arbitrary external momenta. Nevertheless, it agrees with the RWS result in certain regions of momentum space, as illustrated in figure~\ref{fig:RWlimit_AMS_sphere_4pt}. This restricted agreement is expected because the Belavin-Zamolodchikov formula applies only when the Kac labels $r_i$ and $s_i$ satisfy additional conditions.

The Runkel-Watts limit provides a solved corner of the A-series minimal string. The explicit RWS amplitudes may provide useful constraints and insight into the structure of AMS.

\section*{Acknowledgements}
We would like to thank Sasha Artemev, Scott Collier, Rishabh Kaushik, Ben Mazel, and Mykhaylo Usatyuk for useful discussions. 
LE is supported by the European Research Council (ERC) under the European Union’s Horizon 2020 research and innovation programme (grant agreement No 101115511).
VAR is supported by the University of California President’s Postdoctoral Fellowship.
ZW is supported by Heising-Simons Foundation grants \#2024-5307, and by funds from the University of California. The research of MB is funded by the European Union under ERC Synergy Grant MaScAmp (grant agreement No 101167287). We acknowledge use of GPT 5.5 (OpenAI) and Opus 4.8 (Anthropic) for verification purposes.

\appendix

\section{Details of calculations}
\label{app:details-of-calculations}

\paragraph{Derivation of the chamber identity.}
Set $\delta=\frac{q'+q}{2}$ and $x_j=\sqrt{qq'}\,p_j$, and work away from the walls where one of the sine factors below vanishes. The identity to prove \eqref{eq:CLS 3pt limit 2} is
\begin{align}
\sum_{\sigma_2,\sigma_3=\pm1}
(-\sigma_2\sigma_3)
\left\lfloor
\delta+x_1+\sigma_2x_2+\sigma_3x_3
\right\rfloor
=
\left[
\prod_{j=1}^3
\sgn\sin\!\left(2\pi(\delta+x_j)\right)
\right]
\sigma(p_1,p_2,p_3).
\label{eq:appendix-chamber-identity}
\end{align}
To prove this, let us denote $a=\delta+x_1$, $b=x_2$ and $c=x_3$ and set
\begin{align}
N(a,b,c)
&=
\sum_{\sigma_2,\sigma_3=\pm1}
(-\sigma_2\sigma_3)
\left\lfloor a+\sigma_2 b+\sigma_3 c\right\rfloor \ .
\end{align}
We claim that
\begin{align}
N(a,b,c)
=
\sgn\sin(2\pi a)\,
\sgn\sin(2\pi b)\,
\sgn\sin(2\pi c)\,
\mathbf{1}_{\mathcal C}(a,b,c),
\label{eq:universal-chamber-identity}
\end{align}
where $\mathbf{1}_{\mathcal C}(a,b,c)$ is one if
\begin{equation}
\prod_{\sigma_2,\sigma_3=\pm1}
\sin\!\left(\pi(a+\sigma_2 b+\sigma_3 c)\right)<0
\end{equation}
and zero otherwise. \eqref{eq:universal-chamber-identity} implies \eqref{eq:appendix-chamber-identity} because $2\delta=q+q'$ is an integer, so shifting both of the missing sign factors on the RHS of \eqref{eq:universal-chamber-identity} by $\delta$ does not change their product.

Both sides of \eqref{eq:universal-chamber-identity} are periodic under integer shifts of $a,b,c$. They also transform in the same way under the reflections $a\mapsto1-a$, $b\mapsto1-b$, and $c\mapsto1-c$: for example, $N(1-a,b,c)=-N(a,b,c)$, while the chamber indicator is unchanged and $\sgn\sin(2\pi a)$ changes sign. It is therefore enough to check the chamber $0<a,b,c<\frac12$, for which $\sgn \sin(2\pi a)=1$ and similarly for $b$ and $c$. 
In this chamber,
\begin{align}
\lfloor a+b+c\rfloor&=\mathbf{1}_{a+b+c>1},&
\lfloor a+b-c\rfloor&=-\mathbf{1}_{c>a+b},\nonumber\\
\lfloor a-b+c\rfloor&=-\mathbf{1}_{b>a+c},&
\lfloor a-b-c\rfloor&=-\mathbf{1}_{a<b+c}.
\end{align}
Thus both the LHS and the RHS of \eqref{eq:universal-chamber-identity} are piecewise constant and can only jump when $a+\sigma_2 b+\sigma_3c \in \ZZ$. This reduces the equality to checking finitely many cases, which are (i) $a>b+c$ and $a+b+c<1$ and permutations, (ii) $a+b+c>1$ and (iii) $b+c>a$, $a+b>c$, $a+c>b$, $a+b+c<1$. It is then easy to verify \eqref{eq:universal-chamber-identity} for these cases.

\paragraph{The sphere four-point amplitude.}

Let us derive \eqref{eq:RW 04 final} directly from the intersection number formula. For the contact graph, the moduli space is $\bM_{0,4}$ and the only degree-one intersection numbers needed are
\begin{equation}
    \int_{\bM_{0,4}}\psi_i=1\,,
    \qquad
    \int_{\bM_{0,4}}\kappa_1=1\,.
\end{equation}
Hence the vertex integral yields a factor of the VMS quantum volume
\begin{equation}
\mathsf{V}_{0,4}^{(\sqrt{q'/q})}(i\boldsymbol p)
=
\frac{1}{4}\big(\frac{q'}{q}+\frac{q}{q'}\big)-\sum_{i=1}^4p_i^2\,.
\end{equation}
The contact contribution before performing the color sum is
\begin{align}
\mathsf{RW}^{\mathrm{con}}_{0,4}(\boldsymbol p)
&=
\sum_{m=1}^{\infty}
\frac{\mathsf{V}_{0,4}^{(\sqrt{q'/q})}(i\boldsymbol p)}{2\pi^2m^2}
\prod_{i=1}^4\sin(2\pi m x_i)\, .
\label{eq:RW 04 contact color appendix}
\end{align}
Using
\begin{equation}
\prod_{i=1}^4\sin y_i
=
\frac{1}{8}
\sum_{\sigma_2,\sigma_3,\sigma_4=\pm1}
\sigma_2\sigma_3\sigma_4
\cos(y_1+\sigma_2y_2+\sigma_3y_3+\sigma_4y_4)
\end{equation}
and
\begin{equation}
    \sum_{m=1}^\infty \frac{\cos(2\pi m y)}{m^2}
    =\pi^2B_2(\{y\})\,,
\end{equation}
we obtain \eqref{eq:RW 04 contact result}.

It remains to evaluate the exchange graphs. In the $12|34$ channel, the two vertices are copies of $\bM_{0,3}$, and only the degree-zero part of the edge factor contributes. The corresponding term is
\begin{align}
\mathsf{RW}^{\mathrm{ex}}_{0,4;12|34}(\boldsymbol p)
&=
\frac{1}{4\pi^4qq'}
\sum_{m,n\geq 1}
\frac{(-1)^{(q+q')(m+n)}}{mn}
\left(
\frac{\delta_{m\neq n}}{(m-n)^2}
-\frac{1}{(m+n)^2}
\right)
\nonumber\\
&\quad\times
\sin(2\pi m x_1)\sin(2\pi m x_2)
\sin(2\pi n x_3)\sin(2\pi n x_4) \,.
\label{eq:RW 04 exchange color appendix}
\end{align}
We then write
\begin{align}
(-1)^{(q+q')m}\sin(2\pi m x_1)\sin(2\pi m x_2)
&=
\frac{1}{2}
\left[
\cos(2\pi m (x_1-x_2+\tfrac{q+q'}{2}))
\right.
\nonumber\\
&\qquad\left.
-\cos(2\pi m (x_1+x_2+\tfrac{q+q'}{2}))
\right]\,,
\end{align}
and similarly for the pair $(3,4)$. Hence the exchange color sum is reduced to four cosine-cosine sums.

The needed normalized cosine kernel is
\begin{align}
K(u,v)
&=
\frac{1}{\pi^4}
\sum_{m,n\geq1}
\frac{\cos(2\pi m u)\cos(2\pi n v)}{mn}
\left(
\frac{\delta_{m\neq n}}{(m-n)^2}
-\frac{1}{(m+n)^2}
\right).
\label{eq:RW 04 K double sum appendix}
\end{align}
To compute it, first use
\begin{equation}
\frac{1}{mn}
\left(
\frac{\delta_{m\neq n}}{(m-n)^2}
-\frac{1}{(m+n)^2}
\right)
=
\begin{cases}
\dfrac{4}{(m^2-n^2)^2}\,,&m\neq n\,,\\[6pt]
-\dfrac{1}{4m^4}\,,&m=n\,.
\end{cases}
\end{equation}
At fixed $m$, the remaining $n$-sum is given by
\begin{align}
\sum_{\substack{n\geq1\\ n\neq m}}
\frac{\cos(2\pi n v)}{(n^2-m^2)^2}
&=
\cos(2\pi m v)
\left[
\frac{\pi^2B_1(\{v\})^2}{2m^2}
-\frac{\pi^2}{24m^2}
-\frac{3}{16m^4}
\right]
\nonumber\\
&\quad
-\frac{\pi B_1(\{v\})\sin(2\pi m v)}{2m^3}
-\frac{1}{2m^4}\,.
\label{eq:RW fixed m sum appendix}
\end{align}
Substituting \eqref{eq:RW fixed m sum appendix}, the remaining sine sums are evaluated by the Bernoulli Fourier series \eqref{eq:Bernoulli Fourier series sine}, while the cosine sums are evaluated by
\begin{equation} \label{eq:Bernoulli Fourier series cosine}
\sum_{m=1}^\infty\frac{\cos(2\pi mt)}{m^{2k}}
=
(-1)^{k+1}\frac{(2\pi)^{2k}}{2(2k)!}\,
B_{2k}(\{t\})\,,
\quad k\in\mathbb{Z}_{\geq 1}\,.
\end{equation}
This gives precisely the Bernoulli expression \eqref{eq:RW 04 K}. Applying the product-to-sum identity at the two vertices then gives \eqref{eq:RW 04 exchange 1234 result}. The remaining two exchange channels are obtained by permuting the external labels.

\paragraph{Check of the dilaton equation for $\mathsf{RW}_{0,4}$.} 
Set $p_\pm=\frac{q\pm q'}{2\sqrt{qq'}}$, or using our shorthand, $x_\pm=\sqrt{qq'}\,p_\pm=\frac{q\pm q'}{2}$. 
Then $x_+-x_-=q'\in\ZZ$, $2x_\pm\in\ZZ$, and $p_+^2-p_-^2=1$. 
For $(g,n)=(0,3)$, the dilaton equation \eqref{eq:RW dilaton equation} reads
\begin{align}
\partial_p\mathsf{RW}_{0,4}(p_1,p_2,p_3,p)\Big|_{p=p_+}
-
\partial_p\mathsf{RW}_{0,4}(p_1,p_2,p_3,p)\Big|_{p=p_-}
= -\sqrt{qq'}\,\mathsf{RW}_{0,3}(p_1,p_2,p_3) \,.
\end{align}
Let us verify this case, starting from the color-sum representation
of the sphere four-point amplitude \eqref{eq:RW 04 final}. 
The contact term before evaluating the color sum is \eqref{eq:RW 04 contact color appendix},
\begin{align}
\mathsf{RW}^{\mathrm{con}}_{0,4}(\boldsymbol p)
&=
\sum_{m=1}^\infty
\frac{\mathsf{V}_{0,4}^{(\sqrt{q'/q})}(i\boldsymbol p)}{2\pi^2m^2}
\prod_{i=1}^4\sin(2\pi m x_i) \,.
\end{align}
Keeping $p_1,p_2,p_3$ fixed, let us write
\begin{align}
\mathsf{V}_{0,4}^{(\sqrt{q'/q})}(ip_1,ip_2,ip_3,ip)
=\mathsf{V}_3-p^2 \,,
\qquad
\mathsf{V}_3=
\frac14\left(\frac{q'}q+\frac q{q'}\right)
-\sum_{i=1}^3p_i^2 \,.
\end{align}
The part depending on the fourth leg is therefore $(\mathsf{V}_3-p^2)\sin(2\pi m\sqrt{qq'}\,p)$. 
Differentiating before setting $p=p_\pm$, we find
\begin{align}
&\partial_p\!\left[
(\mathsf{V}_3-p^2)\sin(2\pi m\sqrt{qq'}\,p)
\right]
\nonumber\\
&\qquad
=
-2p\sin(2\pi m\sqrt{qq'}\,p)
+2\pi m\sqrt{qq'}(\mathsf{V}_3-p^2)
\cos(2\pi m\sqrt{qq'}\,p) \,.
\end{align}
At $p=p_\pm$, the sine term vanishes and $\cos(2\pi m x_{\pm}) = (-1)^{m(q+q')}$.
Thus the nontrivial identity for the fourth leg is 
\begin{align}
\partial_p & \left[ (\mathsf{V}_3-p^2)\sin(2\pi m\sqrt{qq'}\,p) \right]_{p=p_+}
-
\partial_p \left[ (\mathsf{V}_3-p^2)\sin(2\pi m\sqrt{qq'}\,p) \right]_{p=p_-}
\nonumber\\
& = 2\pi m\sqrt{qq'} \left[ (\mathsf{V}_3-p_+^2)-(\mathsf{V}_3-p_-^2) \right] (-1)^{m(q+q')}
\nonumber\\
& = -2\pi m\sqrt{qq'}\,(-1)^{m(q+q')} \,,
\end{align}
where we used $p_+^2-p_-^2=1$. 
Note that the difference in the square brackets in the second line is nothing but the dilaton equation of the Virasoro minimal string \cite[eq. (4.15a)]{Collier:2023cyw}.
Substituting this identity into the contact graph gives
\begin{align}
&\partial_p\mathsf{RW}^{\mathrm{con}}_{0,4}(p_1,p_2,p_3,p)
\Big|_{p=p_+}
-
\partial_p\mathsf{RW}^{\mathrm{con}}_{0,4}(p_1,p_2,p_3,p)
\Big|_{p=p_-}
\nonumber\\
&\qquad
=
-\sqrt{qq'}
\sum_{m=1}^{\infty} \frac{(-1)^{m(q+q')}}{\pi m} \prod_{j=1}^3\sin(2\pi m x_j)
\nonumber\\
&\qquad
= -\sqrt{qq'}\,\mathsf{RW}_{0,3}(p_1,p_2,p_3) \,,
\end{align}
where in the last line we used the color-sum representation of the three-point amplitude in the first line of \eqref{eq:CLS 3pt limit 1}.

Thus, it remains to show that the exchange graphs do not contribute. 
For example, the $12|34$ channel is \eqref{eq:RW 04 exchange color appendix}. In any channel, if the fourth leg is attached to a vertex of color $r$, the only dependence on $p$ is through $\sin(2\pi r\sqrt{qq'}\,p)$. Hence
\begin{align}
&\partial_p\sin(2\pi r\sqrt{qq'}\,p)\Big|_{p=p_+}
-
\partial_p\sin(2\pi r\sqrt{qq'}\,p)\Big|_{p=p_-}
\nonumber\\
&\qquad
= 2\pi r\sqrt{qq'} \left[ (-1)^{r(q+q')}-(-1)^{r(q-q')} \right] 
= 0 \,.
\end{align}
Therefore only the contact graph survives in the dilaton difference.

\paragraph{From AMS torus one-point amplitude to RWS torus one-point amplitude.}

Take odd labels $r,s$ such that $n=rq'-sq$ is fixed and $r/\mathfrak q\to x\in(0,1)$. From $\mathfrak q'q-\mathfrak q q'=1$, we have
\bal
P_{r,s} = \frac{r\mathfrak q'-s\mathfrak q}{2\sqrt{\mathfrak q\mathfrak q'}} \rightarrow p_1=\frac{n+x}{2\sqrt{qq'}} \, ,
\eal
Since $r$ and $s$ are odd, $n\equiv q'-q\pmod 2$, so
\beq
\left\{\sqrt{qq'}\,p_1+\frac{q+q'}{2}\right\}=\frac{x}{2} \, .
\eeq

We use the identity
\beq
\sum_{\ell\overset{2}{=}1-\mathfrak q'+s}^{\mathfrak q'-s-1}
|\ell+y|
=
\frac{(\mathfrak q'-s)^2+y^2}{2}-\frac{1}{6}
-2B_2\left(
\left\{\frac{\mathfrak q'-s-1-y}{2}\right\}
\right),
\qquad |y|<\mathfrak q'-s \, .
\eeq
Applying it to the inner sum in \eqref{eq:AMS-torus-one-point}, together with
\beq
\sum_{m\overset{2}{=}1-\mathfrak q+r}^{\mathfrak q-r-1}m^2
=\frac{1}{3}(\mathfrak q-r)((\mathfrak q-r)^2-1) \, ,
\eeq
gives
\bal
T_{r,s}^{(\mathfrak q,\mathfrak q')}
={}&
\frac{\mathfrak q-r}{2\mathfrak q}
\left[
\mathfrak q^2+\mathfrak q'^2
-(r\mathfrak q'-s\mathfrak q)^2
\right]
\nonumber\\
&\quad
+6\mathfrak q
\sum_{m\overset{2}{=}1-\mathfrak q+r}^{\mathfrak q-r-1}
B_2\left(
\left\{
\frac{\mathfrak q'-s-1}{2}
-\frac{m\mathfrak q'}{2\mathfrak q}
\right\}
\right)
+\mathcal O(\mathfrak q) \, .
\eal
The error comes from at most finitely many endpoint terms and vanishes after division by $\mathfrak q\mathfrak q'$.

After division by $\mathfrak q\mathfrak q'$, the first term yields a polynomial. For the second term, we insert
\beq
B_2(\{u\})=\frac{1}{2\pi^2}
\sum_{\ell\neq0}\frac{e^{2\pi i\ell u}}{\ell^2} \, .
\eeq
After performing the finite sum over $m$, 
\bal
\sum_{m\overset{2}{=}1-\mathfrak q+r}^{\mathfrak q-r-1}
B_2\left(
\left\{
\frac{\mathfrak q'-s-1}{2}
-\frac{m\mathfrak q'}{2\mathfrak q}
\right\}
\right)=\frac{1}{2\pi^2} \sum_{\ell\neq 0} \frac{e^{i\pi \ell(\mathfrak q'-s-1)}}{\ell^2} \frac{\sin(\pi \ell (\mathfrak q-r) \frac{\mathfrak q'}{\mathfrak q})}{\sin(\pi \ell \frac{\mathfrak q'}{\mathfrak q})}\, ,
\eal
only the modes $\ell=qk$ survive in the limit. Using
\beq
e^{\pi iqk(\mathfrak q'-s-1)}
\frac{
\sin\left(\pi qk(\mathfrak q-r)\frac{\mathfrak q'}{\mathfrak q}\right)
}{
\sin\left(\pi qk\frac{\mathfrak q'}{\mathfrak q}\right)
}
=
-\frac{\sin\frac{\pi kr}{\mathfrak q}}{\sin\frac{\pi k}{\mathfrak q}} \, ,
\eeq
we obtain
\bal
\lim_{\mathfrak q,\mathfrak q'\to\infty}
\frac{1}{\mathfrak q'}
&\sum_{m\overset{2}{=}1-\mathfrak q+r}^{\mathfrak q-r-1}
B_2\left(
\left\{
\frac{\mathfrak q'-s-1}{2}
-\frac{m\mathfrak q'}{2\mathfrak q}
\right\}
\right)
\nonumber\\
&=
-\frac{1}{2\pi^3qq'}
\sum_{k\neq0}\frac{\sin(\pi kx)}{k^3}
=
-\frac{2}{3qq'}B_3\left(\frac{x}{2}\right) \, .
\eal
Consequently,
\beq
\lim_{\mathfrak q,\mathfrak q'\to\infty}
\frac{T_{r,s}^{(\mathfrak q,\mathfrak q')}}{192\mathfrak q\mathfrak q'}
=
\frac{1-x}{384}
\left(
\frac{q'}{q}+\frac{q}{q'}-4p_1^2
\right)
-\frac{1}{48qq'}B_3\left(\frac{x}{2}\right) \, .
\eeq
Using $B_1(x/2)=(x-1)/2$ and
\beq
V_{1,1}^{(\sqrt{q'/q})}(ip_1)
=
\frac{1}{24}
\left[
\frac{1}{4}\left(\frac{q'}{q}+\frac{q}{q'}\right)-p_1^2
\right] \, ,
\eeq
this is precisely
\beq
\lim_{\mathfrak q,\mathfrak q'\to\infty}
\frac{T_{r,s}^{(\mathfrak q,\mathfrak q')}}{192\mathfrak q\mathfrak q'}
=
\mathsf{RW}_{1,1}(p_1) \, ,
\eeq
as claimed.

\bibliographystyle{JHEP}
\bibliography{bib}

\providecommand{\href}[2]{#2}\begingroup\raggedright\begin{thebibliography}{10}

\bibitem{Brezin:1990rb}
E.~Brezin and V.~Kazakov, \emph{{Exactly Solvable Field Theories of Closed
  Strings}}, \href{https://doi.org/10.1016/0370-2693(90)90818-Q}{\emph{Phys.
  Lett. B} {\bfseries 236} (1990) 144}.

\bibitem{Douglas:1989ve}
M.~R. Douglas and S.~H. Shenker, \emph{{Strings in Less Than One-Dimension}},
  \href{https://doi.org/10.1016/0550-3213(90)90522-F}{\emph{Nucl. Phys. B}
  {\bfseries 335} (1990) 635}.

\bibitem{Gross:1989vs}
D.~J. Gross and A.~A. Migdal, \emph{{Nonperturbative Two-Dimensional Quantum
  Gravity}}, \href{https://doi.org/10.1103/PhysRevLett.64.127}{\emph{Phys. Rev.
  Lett.} {\bfseries 64} (1990) 127}.

\bibitem{Seiberg:2004at}
N.~Seiberg and D.~Shih, \emph{{Minimal string theory}},
  \href{https://doi.org/10.1016/j.crhy.2004.12.007}{\emph{Comptes Rendus
  Physique} {\bfseries 6} (2005) 165}
  [\href{https://arxiv.org/abs/hep-th/0409306}{{\ttfamily hep-th/0409306}}].

\bibitem{Klebanov:1991qa}
I.~R. Klebanov, \emph{{String theory in two-dimensions}},  in \emph{{Spring
  School on String Theory and Quantum Gravity (to be followed by Workshop)}},
  7, 1991, \href{https://arxiv.org/abs/hep-th/9108019}{{\ttfamily
  hep-th/9108019}}.

\bibitem{Ginsparg:1993is}
P.~H. Ginsparg and G.~W. Moore, \emph{{Lectures on 2-D gravity and 2-D string
  theory}},  in \emph{{Theoretical Advanced Study Institute (TASI 92): From
  Black Holes and Strings to Particles}}, pp.~277--469, 10, 1993,
  \href{https://arxiv.org/abs/hep-th/9304011}{{\ttfamily hep-th/9304011}}.

\bibitem{Jevicki:1993qn}
A.~Jevicki, \emph{{Development in 2-d string theory}},  in \emph{{Workshop on
  String Theory, Gauge Theory and Quantum Gravity}}, 9, 1993,
  \href{https://arxiv.org/abs/hep-th/9309115}{{\ttfamily hep-th/9309115}},
  \href{https://doi.org/10.1142/9789814447072_0004}{DOI}.

\bibitem{Polchinski:1994mb}
J.~Polchinski, \emph{{What is string theory?}},  in \emph{{NATO Advanced Study
  Institute: Les Houches Summer School, Session 62: Fluctuating Geometries in
  Statistical Mechanics and Field Theory}}, 11, 1994,
  \href{https://arxiv.org/abs/hep-th/9411028}{{\ttfamily hep-th/9411028}}.

\bibitem{Balthazar:2019rnh}
B.~Balthazar, V.~A. Rodriguez and X.~Yin, \emph{{ZZ instantons and the
  non-perturbative dual of c = 1 string theory}},
  \href{https://doi.org/10.1007/JHEP05(2023)048}{\emph{JHEP} {\bfseries 05}
  (2023) 048} [\href{https://arxiv.org/abs/1907.07688}{{\ttfamily
  1907.07688}}].

\bibitem{Collier:2023cyw}
S.~Collier, L.~Eberhardt, B.~M{\"u}hlmann and V.~A. Rodriguez, \emph{{The
  Virasoro minimal string}},
  \href{https://doi.org/10.21468/SciPostPhys.16.2.057}{\emph{SciPost Phys.}
  {\bfseries 16} (2024) 057}
  [\href{https://arxiv.org/abs/2309.10846}{{\ttfamily 2309.10846}}].

\bibitem{Saad:2019lba}
P.~Saad, S.~H. Shenker and D.~Stanford, \emph{{JT gravity as a matrix
  integral}},  \href{https://arxiv.org/abs/1903.11115}{{\ttfamily 1903.11115}}.

\bibitem{Collier:2024kmo}
S.~Collier, L.~Eberhardt, B.~M{\"u}hlmann and V.~A. Rodriguez, \emph{{Complex
  Liouville String}}, \href{https://doi.org/10.1103/k74n-s63l}{\emph{Phys. Rev.
  Lett.} {\bfseries 134} (2025) 251602}
  [\href{https://arxiv.org/abs/2409.17246}{{\ttfamily 2409.17246}}].

\bibitem{Collier:2024kwt}
S.~Collier, L.~Eberhardt, B.~M{\"u}hlmann and V.~A. Rodriguez, \emph{{The
  complex Liouville string: The worldsheet}},
  \href{https://doi.org/10.21468/SciPostPhys.19.2.033}{\emph{SciPost Phys.}
  {\bfseries 19} (2025) 033}
  [\href{https://arxiv.org/abs/2409.18759}{{\ttfamily 2409.18759}}].

\bibitem{Collier:2024lys}
S.~Collier, L.~Eberhardt, B.~M{\"u}hlmann and V.~A. Rodriguez, \emph{{The
  complex Liouville string: The matrix integral}},
  \href{https://doi.org/10.21468/SciPostPhys.18.5.154}{\emph{SciPost Phys.}
  {\bfseries 18} (2025) 154}
  [\href{https://arxiv.org/abs/2410.07345}{{\ttfamily 2410.07345}}].

\bibitem{Collier:2024mlg}
S.~Collier, L.~Eberhardt, B.~M{\"u}hlmann and V.~A. Rodriguez, \emph{{The
  complex Liouville string: Worldsheet boundaries and non-perturbative
  effects}}, \href{https://doi.org/10.21468/SciPostPhys.19.2.034}{\emph{SciPost
  Phys.} {\bfseries 19} (2025) 034}
  [\href{https://arxiv.org/abs/2410.09179}{{\ttfamily 2410.09179}}].

\bibitem{Collier:2025pbm}
S.~Collier, L.~Eberhardt and B.~M{\"u}hlmann, \emph{{The complex Liouville
  string: The gravitational path integral}},
  \href{https://doi.org/10.21468/SciPostPhys.19.4.115}{\emph{SciPost Phys.}
  {\bfseries 19} (2025) 115}
  [\href{https://arxiv.org/abs/2501.10265}{{\ttfamily 2501.10265}}].

\bibitem{Collier:2025lux}
S.~Collier, L.~Eberhardt and B.~M{\"u}hlmann, \emph{{A microscopic realization
  of dS$_3$}},
  \href{https://doi.org/10.21468/SciPostPhys.18.4.131}{\emph{SciPost Phys.}
  {\bfseries 18} (2025) 131}
  [\href{https://arxiv.org/abs/2501.01486}{{\ttfamily 2501.01486}}].

\bibitem{Collier:2026pxi}
S.~Collier, L.~Eberhardt and V.~A. Rodriguez, \emph{{$c=1$ strings as a matrix
  integral}},  \href{https://arxiv.org/abs/2604.06301}{{\ttfamily 2604.06301}}.

\bibitem{Runkel:2001ng}
I.~Runkel and G.~M.~T. Watts, \emph{{A Nonrational CFT with c = 1 as a limit of
  minimal models}},
  \href{https://doi.org/10.1088/1126-6708/2001/09/006}{\emph{JHEP} {\bfseries
  09} (2001) 006} [\href{https://arxiv.org/abs/hep-th/0107118}{{\ttfamily
  hep-th/0107118}}].

\bibitem{Schomerus:2003vv}
V.~Schomerus, \emph{{Rolling tachyons from Liouville theory}},
  \href{https://doi.org/10.1088/1126-6708/2003/11/043}{\emph{JHEP} {\bfseries
  11} (2003) 043} [\href{https://arxiv.org/abs/hep-th/0306026}{{\ttfamily
  hep-th/0306026}}].

\bibitem{McElgin:2007ak}
W.~McElgin, \emph{{Notes on Liouville Theory at c {\ensuremath{<}}= 1}},
  \href{https://doi.org/10.1103/PhysRevD.77.066009}{\emph{Phys. Rev. D}
  {\bfseries 77} (2008) 066009}
  [\href{https://arxiv.org/abs/0706.0365}{{\ttfamily 0706.0365}}].

\bibitem{Sen:2024nfd}
A.~Sen and B.~Zwiebach, \emph{{String Field Theory: A Review}},
  \href{https://arxiv.org/abs/2405.19421}{{\ttfamily 2405.19421}}.

\bibitem{Eynard:2007kz}
B.~Eynard and N.~Orantin, \emph{{Invariants of algebraic curves and topological
  expansion}}, \href{https://doi.org/10.4310/CNTP.2007.v1.n2.a4}{\emph{Commun.
  Num. Theor. Phys.} {\bfseries 1} (2007) 347}
  [\href{https://arxiv.org/abs/math-ph/0702045}{{\ttfamily math-ph/0702045}}].

\bibitem{Chekhov:2006vd}
L.~Chekhov, B.~Eynard and N.~Orantin, \emph{{Free energy topological expansion
  for the 2-matrix model}},
  \href{https://doi.org/10.1088/1126-6708/2006/12/053}{\emph{JHEP} {\bfseries
  12} (2006) 053} [\href{https://arxiv.org/abs/math-ph/0603003}{{\ttfamily
  math-ph/0603003}}].

\bibitem{Artemev:2025pvk}
A.~Artemev, \emph{{x-y swap for a (2, 2p+1) minimal string}},
  \href{https://doi.org/10.1103/pd2y-j2x5}{\emph{Phys. Rev. D} {\bfseries 112}
  (2025) 046019} [\href{https://arxiv.org/abs/2506.09222}{{\ttfamily
  2506.09222}}].

\bibitem{Dorn:1994xn}
H.~Dorn and H.~J. Otto, \emph{{Two and three point functions in Liouville
  theory}}, \href{https://doi.org/10.1016/0550-3213(94)00352-1}{\emph{Nucl.
  Phys.} {\bfseries B429} (1994) 375}
  [\href{https://arxiv.org/abs/hep-th/9403141}{{\ttfamily hep-th/9403141}}].

\bibitem{Zamolodchikov:1995aa}
A.~B. Zamolodchikov and A.~B. Zamolodchikov, \emph{{Structure constants and
  conformal bootstrap in Liouville field theory}},
  \href{https://doi.org/10.1016/0550-3213(96)00351-3}{\emph{Nucl. Phys.}
  {\bfseries B477} (1996) 577}
  [\href{https://arxiv.org/abs/hep-th/9506136}{{\ttfamily hep-th/9506136}}].

\bibitem{Teschner:1995yf}
J.~Teschner, \emph{{On the Liouville three point function}},
  \href{https://doi.org/10.1016/0370-2693(95)01200-A}{\emph{Phys. Lett. B}
  {\bfseries 363} (1995) 65}
  [\href{https://arxiv.org/abs/hep-th/9507109}{{\ttfamily hep-th/9507109}}].

\bibitem{Eberhardt:2023mrq}
L.~Eberhardt, \emph{{Notes on crossing transformations of Virasoro conformal
  blocks}},  \href{https://arxiv.org/abs/2309.11540}{{\ttfamily 2309.11540}}.

\bibitem{Ribault:2015sxa}
S.~Ribault and R.~Santachiara, \emph{{Liouville theory with a central charge
  less than one}}, \href{https://doi.org/10.1007/JHEP08(2015)109}{\emph{JHEP}
  {\bfseries 08} (2015) 109}
  [\href{https://arxiv.org/abs/1503.02067}{{\ttfamily 1503.02067}}].

\bibitem{Rodriguez:2025rte}
V.~A. Rodriguez, M.~Usatyuk and Z.-Y. Wang, \emph{{ADE Minimal Strings and
  Multi-Matrix Duals}},  \href{https://arxiv.org/abs/2511.21851}{{\ttfamily
  2511.21851}}.

\bibitem{Khromov:2025awh}
D.~Khromov and A.~Litvinov, \emph{{On correlation numbers V$_{0,4}$ and
  V$_{1,1}$ in Virasoro minimal string theory}},
  \href{https://doi.org/10.1007/JHEP04(2026)126}{\emph{JHEP} {\bfseries 04}
  (2026) 126} [\href{https://arxiv.org/abs/2509.25960}{{\ttfamily
  2509.25960}}].

\bibitem{Eynard:2011ga}
B.~Eynard, \emph{{Invariants of spectral curves and intersection theory of
  moduli spaces of complex curves}},
  \href{https://doi.org/10.4310/CNTP.2014.v8.n3.a4}{\emph{Commun. Num. Theor.
  Phys.} {\bfseries 8} (2014) 541}
  [\href{https://arxiv.org/abs/1110.2949}{{\ttfamily 1110.2949}}].

\bibitem{Dunin-Barkowski:2012kbi}
P.~Dunin-Barkowski, N.~Orantin, S.~Shadrin and L.~Spitz, \emph{{Identification
  of the Givental formula with the spectral curve topological recursion
  procedure}}, \href{https://doi.org/10.1007/s00220-014-1887-2}{\emph{Commun.
  Math. Phys.} {\bfseries 328} (2014) 669}
  [\href{https://arxiv.org/abs/1211.4021}{{\ttfamily 1211.4021}}].

\bibitem{Mazel:2024alu}
B.~Mazel, J.~Sandor, C.~Wang and X.~Yin, \emph{{Conformal Perturbation Theory
  and Tachyon-Dilaton Eschatology via String Fields}},
  \href{https://arxiv.org/abs/2403.14544}{{\ttfamily 2403.14544}}.

\bibitem{Rodriguez:2023kkl}
V.~A. Rodriguez, \emph{{A two-dimensional string cosmology}},
  \href{https://doi.org/10.1007/JHEP06(2023)161}{\emph{JHEP} {\bfseries 06}
  (2023) 161} [\href{https://arxiv.org/abs/2302.06625}{{\ttfamily
  2302.06625}}].

\bibitem{Rodriguez:2023wun}
V.~A. Rodriguez, \emph{{The torus one-point diagram in two-dimensional string
  cosmology}}, \href{https://doi.org/10.1007/JHEP07(2023)050}{\emph{JHEP}
  {\bfseries 07} (2023) 050}
  [\href{https://arxiv.org/abs/2304.13043}{{\ttfamily 2304.13043}}].

\bibitem{Gaberdiel:2011aa}
M.~R. Gaberdiel and P.~Suchanek, \emph{{Limits of Minimal Models and Continuous
  Orbifolds}}, \href{https://doi.org/10.1007/JHEP03(2012)104}{\emph{JHEP}
  {\bfseries 03} (2012) 104} [\href{https://arxiv.org/abs/1112.1708}{{\ttfamily
  1112.1708}}].

\bibitem{Witten:1991we}
E.~Witten, \emph{{On quantum gauge theories in two-dimensions}},
  \href{https://doi.org/10.1007/BF02100009}{\emph{Commun. Math. Phys.}
  {\bfseries 141} (1991) 153}.

\bibitem{Witten:1992xu}
E.~Witten, \emph{{Two-dimensional gauge theories revisited}},
  \href{https://doi.org/10.1016/0393-0440(92)90034-X}{\emph{J. Geom. Phys.}
  {\bfseries 9} (1992) 303}
  [\href{https://arxiv.org/abs/hep-th/9204083}{{\ttfamily hep-th/9204083}}].

\bibitem{Blau:1993hj}
M.~Blau and G.~Thompson, \emph{{Lectures on 2-d gauge theories: Topological
  aspects and path integral techniques}},
  \href{https://arxiv.org/abs/hep-th/9310144}{{\ttfamily hep-th/9310144}}.

\bibitem{Cordes:1994fc}
S.~Cordes, G.~W. Moore and S.~Ramgoolam, \emph{{Lectures on 2-d Yang-Mills
  theory, equivariant cohomology and topological field theories}},
  \href{https://doi.org/10.1016/0920-5632(95)00434-B}{\emph{Nucl. Phys. Proc.
  Suppl.} {\bfseries 41} (1995) 184}
  [\href{https://arxiv.org/abs/hep-th/9411210}{{\ttfamily hep-th/9411210}}].

\bibitem{Giribet:2009hm}
G.~Giribet, \emph{{On AGT description of N=2 SCFT with N(f) = 4}},
  \href{https://doi.org/10.1007/JHEP01(2010)097}{\emph{JHEP} {\bfseries 01}
  (2010) 097} [\href{https://arxiv.org/abs/0912.1930}{{\ttfamily 0912.1930}}].

\bibitem{Alday:2009aq}
L.~F. Alday, D.~Gaiotto and Y.~Tachikawa, \emph{{Liouville Correlation
  Functions from Four-dimensional Gauge Theories}},
  \href{https://doi.org/10.1007/s11005-010-0369-5}{\emph{Lett. Math. Phys.}
  {\bfseries 91} (2010) 167} [\href{https://arxiv.org/abs/0906.3219}{{\ttfamily
  0906.3219}}].

\bibitem{Gaiotto:2009we}
D.~Gaiotto, \emph{{N=2 dualities}},
  \href{https://doi.org/10.1007/JHEP08(2012)034}{\emph{JHEP} {\bfseries 08}
  (2012) 034} [\href{https://arxiv.org/abs/0904.2715}{{\ttfamily 0904.2715}}].

\bibitem{Eynard:2015aea}
B.~Eynard, T.~Kimura and S.~Ribault, \emph{{Random matrices}},
  \href{https://arxiv.org/abs/1510.04430}{{\ttfamily 1510.04430}}.

\bibitem{Ribault:2014hia}
S.~Ribault, \emph{{Conformal field theory on the plane}},
  \href{https://arxiv.org/abs/1406.4290}{{\ttfamily 1406.4290}}.

\bibitem{Roussillon:2025tmv}
J.~Roussillon and I.~Tsiares, \emph{{On the Virasoro Crossing Kernels at
  Rational Central Charge}},
  \href{https://doi.org/10.21468/SciPostPhys.20.6.167}{\emph{SciPost Phys.}
  {\bfseries 20} (2026) 167}
  [\href{https://arxiv.org/abs/2512.03172}{{\ttfamily 2512.03172}}].

\bibitem{Ribault:2024rvk}
S.~Ribault, \emph{{Exactly solvable conformal field theories}},
  \href{https://arxiv.org/abs/2411.17262}{{\ttfamily 2411.17262}}.

\bibitem{Belavin:2006ex}
A.~A. Belavin and A.~B. Zamolodchikov, \emph{{Integrals over moduli spaces,
  ground ring, and four-point function in minimal Liouville gravity}},
  \href{https://doi.org/10.1007/s11232-006-0075-8}{\emph{Theor. Math. Phys.}
  {\bfseries 147} (2006) 729}.

\end{thebibliography}\endgroup
\end{document}